\documentclass{article}
\usepackage{arxiv}
\usepackage[utf8]{inputenc} 
\usepackage[T1]{fontenc}    
\usepackage[colorlinks,urlcolor=blue,citecolor=blue]{hyperref}
\usepackage{url}            
\usepackage{booktabs}       
\usepackage{amsfonts}       
\usepackage{nicefrac}       
\usepackage{microtype}      
\usepackage{lipsum}		
\usepackage{graphicx}
\usepackage{subcaption}
\usepackage{float}
\usepackage{doi}
\usepackage[style=numeric]{biblatex}  
\usepackage{amsmath}
\usepackage{mathtools}
\usepackage{xcolor}
\usepackage{comment}

\title{Bridging statistical mechanics and thermodynamics away from equilibrium: a data-driven approach for learning internal variables and their dynamics}

\author{ \hspace{1mm}Weilun Qiu\\
	Department of Mechanical Engineering \\
    and Applied Mechanics\\
	University of Pennsylvania\\
	Philadelphia, PA 19104 \\
	\texttt{wlqiu@seas.upenn.edu} \\
    \And
    \hspace{1mm}Shenglin Huang\\
	Department of Mechanical Engineering \\
    and Applied Mechanics\\
	University of Pennsylvania\\
	Philadelphia, PA 19104 \\
	\texttt{simonhuang1993@126.com} \\
    \And
    \hspace{1mm}Celia Reina\\
	Department of Mechanical Engineering \\
    and Applied Mechanics\\
	University of Pennsylvania\\
	Philadelphia, PA 19104 \\
	\texttt{creina@seas.upenn.edu} \\
}

\date{}

\renewcommand{\headeright}{}
\renewcommand{\undertitle}{}

\begin{document}
\maketitle

\begin{abstract}
    Thermodynamics with internal variables is a common approach in continuum mechanics to model inelastic (i.e., non-equilibrium) material behavior. It consists of enlarging the space of the state variables 
    by introducing internal variables to eliminate the memory effects that would otherwise arise in the constitutive response when driving the system away from equilibrium. While this approach is computationally and theoretically very attractive, it currently lacks a well-established statistical mechanics foundation. As a result, internal variables are typically chosen phenomenologically and lack a direct link to the underlying atomistic or particle description. This hinders the predictability of the ensuing continuum models as well as the inverse problem of material design. In this work, we propose a machine learning approach that directly tackles these underlying issues, by learning internal variables and the evolution equations of the system, consistently with the principles of statistical mechanics and thermodynamics. The proposed approach leverages the following machine learning techniques (i) the information bottleneck (IB) method to ensure that the learned internal variables are functions of the microstates and are capable of capturing the salient feature of the microscopic distribution; (ii) conditional normalizing flows to represent arbitrary probability distributions of the microscopic states as functions of the state variables (these will be distinct from the Boltzmann distribution away from equilibrium); and (iii) Variational Onsager Neural Networks (VONNs) to guarantee thermodynamic consistency of the learned evolution equations and that the state variables are sufficient to predict the future state of the system in the absence of memory effects. The resulting framework, called IB-VONNs, is here tested on two problems on colloidal systems, governed at the microscale by overdamped Langevin dynamics. The first one is a prototypical model for a colloidal particle in an optical trap, which can be solved analytically thanks to its simplicity, and it is thus ideal to verify the framework. The second problem is a one-dimensional phase-transforming system, whose macroscopic description still lacks a statistical mechanics foundation under general conditions. The results in both cases indicate that the proposed machine learning strategy can indeed bridge statistical mechanics and thermodynamics with internal variables away from equilibrium.
\end{abstract}

\keywords{Non-equilibrium thermodynamics, inelasticity, internal variables, continuum mechanics, information bottleneck, conditional normalizing flows, Variational Onsager Neural Networks, overdamped Langevin dynamics}

\section{Introduction} \label{sec: intro}
A fundamental scientific and technological challenge in mechanics of materials is to understand, predict, and control inelastic (i.e., non-equilibrium) material behavior: from crystal plasticity in metallic systems \cite{lubliner2008plasticity}, to solid-solid phase transformations  \cite{levitas1998thermomechanical}, 
to particle rearrangements in disordered media 
\cite{schoenholz2017relationship}.
Unlike in the equilibrium regime \cite{pathria2017statistical}, where we have a well-established statistical mechanics foundation for macroscopic thermodynamic models, there is no universally accepted theoretical framework for systems and processes that lie away from equilibrium. 


Historically, two approaches have been widely adopted to model inelastic material phenomena, which differ in their choice of macroscopic state variables to describe the system. The first approach 
uses the same set of macroscopic state variables, $\boldsymbol{\chi}$, as in equilibrium thermodynamics (i.e., elasticity), e.g., the number of particles, temperature, and the deformation gradient in the canonical ensemble. The main difficulty of this approach is that, in general, the mechanical response becomes a functional of the entire history of the state variables when the system is driven away from equilibrium \cite{coleman1960approximation,coleman1961foundations}. This is in contrast to the equilibrium regime (i.e., elasticity), where the stress is a function of their instantaneous value.
The second approach is that of thermodynamics with internal variables \cite{coleman1967thermodynamics,lubliner1972thermodynamic,maugin1994thermodynamics,ottinger2005beyond}. The idea of this formalism is to introduce additional variables $\alpha$, called internal variables, to fully characterize the state of system without memory effects. That is, the dependent variables become direct functions of the enlarged state space ($\boldsymbol{\chi}$, $\boldsymbol{\alpha}$) and the evolution equations become Markovian. The lack of memory effects is in general highly convenient from both a modeling and computational standpoint, as it significantly reduces the complexity of the constitutive relations as well as the computer memory needed in numerical simulations. Furthermore, they provide, at least in theory, a more direct pathway to understand the vividly sought-after structure-property relations, essential for material design. These advantages have made thermodynamics with internal variables widely used in mechanics of materials, where it has been highly successful to model inelastic phenomena, such as 
crystalline plasticity \cite{rice1971inelastic}, 
viscoelasticity \cite{holzapfel1996large}, 
damage mechanics, \cite{mazars1989continuum}
and mechanics of granular materials 
\cite{tengattini2014thermomechanical}. 

Internal variables are physically thought of as microstructural descriptors (although they are often chosen phenomenologically) and they are typically considered to be measurable, but not controllable \cite{maugin1999thermomechanics,maugin1994thermodynamics, ottinger2005beyond} (i.e., their value cannot be directly adjusted to prescribed values through the action of external forces). In equilibrium, the internal variables reduce to functions of the usual state variables, i.e., $\boldsymbol{\alpha}=\boldsymbol{\alpha}(\boldsymbol{\chi})$, and they are thus not necessary to fully describe the state of the system \cite{muschik1991internal}. 
Away from equilibrium, internal variables are independent variables, and additional evolution laws for such variables are thus needed to complete the governing equations. 

Concepts similar to internal variables have also been adopted in other fields. One such concept is that of order parameter \cite{bagchi2018statistical,maugin1994thermodynamics2, sethna2021statistical}, widely used in the study of phase transitions. Similarly to thermodynamics with internal variables, when a system is outside of equilibrium, the order parameter is a free variable; while in equilibrium, the order parameter becomes a function of the other state variables. Another related concept in the field of biomolecular simulations is that of collective variables (CVs) \cite{gkeka2020machine}, or reaction coordinates (RCs) \cite{peters2016reaction}. The long-time dynamical behavior of molecular systems is typically governed by a small number of CVs. A good choice of CVs should also bring in no memory effects \cite{best2005reaction}. 

Despite the appeal and success of thermodynamics with internal variables, there is currently no automatic strategy to identify a suitable choice of internal variables \cite{van2020roadmap}, at least not a widely accepted approach. Consequently, modeler input and physical intuition end up guiding most modeling efforts, which undoubtedly leads to subjectivity in the resulting models. The shortcomings of this "intuition-based" approach are well illustrated by one of the most common choices of macroscopic variables to describe inelastic material behavior (e.g., plasticity,  viscoelasticity, phase transformation or growth). These variables often include the temperature, the deformation gradient tensor $\boldsymbol{F}$ characterizing the local shape change, and internal variables $\boldsymbol{F}^i$ describing the inelastic contribution to the total deformation. 
Traditionally, these are assumed to follow a kinematic relation of the form $\boldsymbol{F} = \boldsymbol{F}^e \boldsymbol{F}^i$ \cite{ LeeLiu1967,Rice1971,Rodriguez1994,kamrin2010nonlinear,govindjee1997presentation,levitas1998thermomechanical,shanthraj2017elasto},
with $\boldsymbol{F}^e$ being the elastic deformation tensor. Despite the widespread use of this kinematic characterization, recent studies indicate that, while its validity and rigorous link to the microstructure are justified in the context of crystal plasticity induced by dislocation glide \cite{reina2016derivation,reina2017incompressible,reina2018kinematics}, it is, in general, a continuum approximation with unknown accuracy \cite{reina2017incompressible,reina2018kinematics}. This is, of course, undesirable per se, but in addition, it forces the modeler to use phenomenological evolution equations. Indeed, if the macroscopic variables are not a direct function of microscopic quantities, it is impossible to derive the macroscopic dynamics from the microscopic evolution equations. This often leads to a lack of predictability of the ensuing continuum models and hinders the inverse problem of material design.

These examples highlight, on the one hand, the limitation of using physical intuition to come up with appropriate macroscopic variables, and, on the other hand, the lack of model transferability between inelastic phenomena in mechanics. An automatic strategy to identify good macroscopic variables from the microscopic description is therefore a crucial need, as it would broadly enable predictive simulations, and would greatly accelerate the timeline for model development.

The problem of automatic feature extraction is actually widely studied in machine learning 
\cite{murphy2012machine,goodfellow2016deep}, 
and it is thus no surprise that such techniques are being adopted to find coarse-grained descriptors. Common approaches of feature extraction include principal component analysis (PCA) \cite{jolliffe1986principal,jolliffe2016principal,ryckelynck2009hyper}, diffusion maps \cite{coifman2006diffusion,coifman2008diffusion,evans2023computing,rohrdanz2011determination} and autoencoders \cite{gonzalez2018deep,lee2020model,varolgunecs2020interpretable,hernandez2018variational,wehmeyer2018time}. 
We refer the reader to \cite{gkeka2020machine} and references therein for a review on macroscopic variable identification from molecular dynamics data. However, while their adoption in biochemistry, for instance, is widespread, their adoption in mechanics to identify internal state variables of the type described above is only emerging \cite{liu2023learning,masi2022multiscale,masi2023evolution}. 
Existing efforts are primarily concerned with coarse-graining continuum models and develop homogenized models with internal variables. 
In contrast, the focus of the present investigation is in the probabilistic coarse-graining of atomistic/particle models in the spirit of non-equilibrium statistical mechanics and thermodynamics with internal variables, analogously to how the Boltzmann distribution describes the state of the system in the canonical ensemble (the precise connection between the density of states and the internal variables away from equilibrium will be discussed next). Although we are not aware of any data-driven approach of this kind, it is worth noting that the structural variable (relating to the radial distribution function) that locally correlates with rearrangements in disordered media has been an open challenge for 50 years, puzzled by the fact that systems with very similar structure can exhibit vastly different dynamics, and it may have recently been solved thanks to machine learning \cite{cubuk2015identifying}. The approach there used to discover such internal variable is specific to rearranging particle systems, but this success story exemplifies the untapped potential that machine learning techniques can have in guiding internal variable identification and, more broadly, in bridging statistical mechanics and thermodynamics away from equilibrium. 

\begin{figure}[t]
    \centering
    \includegraphics[width=0.85\textwidth]{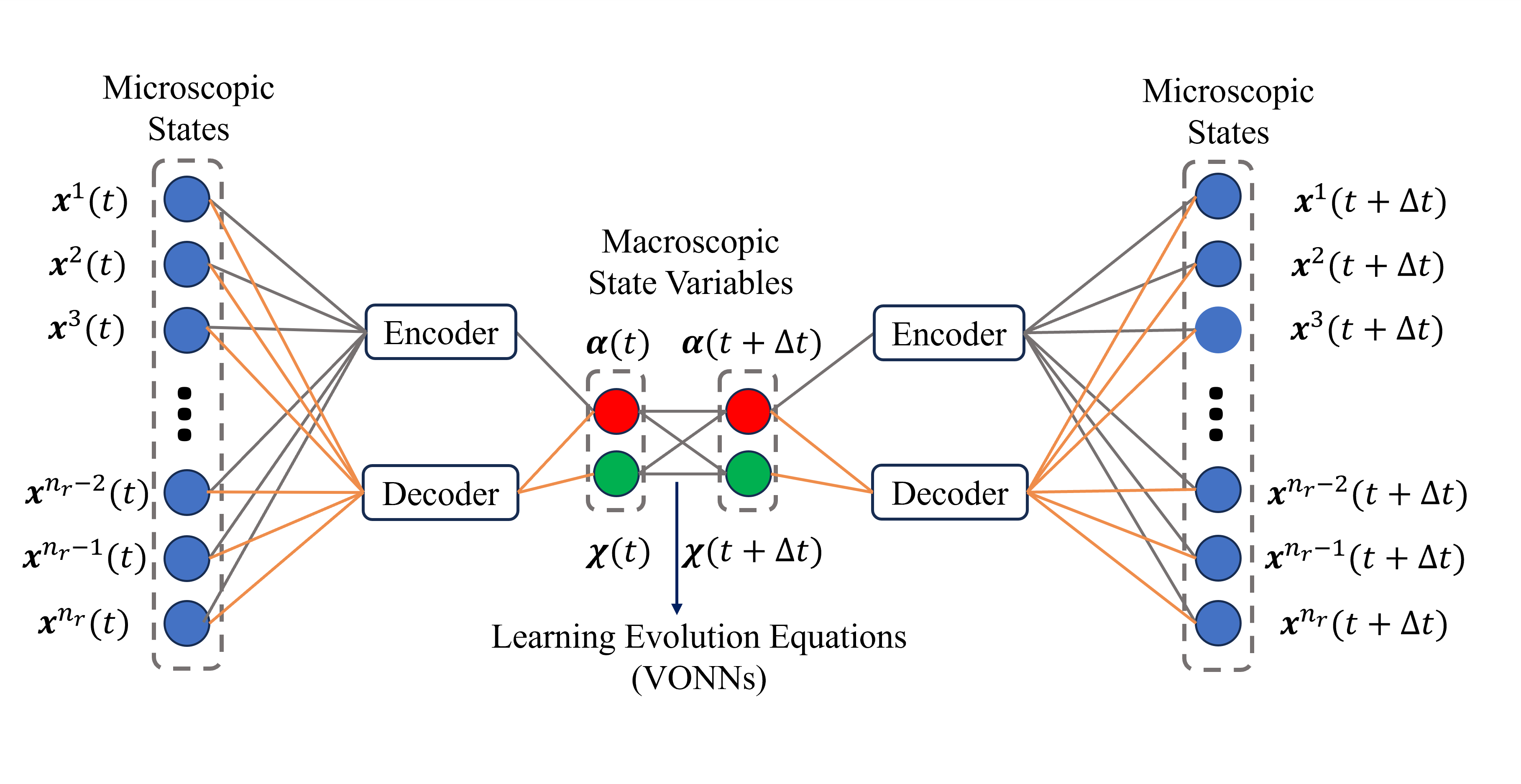}
    \caption{Overview of the IB-VONNs framework. It is composed of an encoder-decoder architecture that relates the microscopic probability distribution with the state variables at all times (based on the Information Bottleneck principle), and Variational Onsager Neural Networks to discover the evolution equations of the macroscopic fields. The state variables are composed of the usual variables to describe the system in equilibrium $\chi$ and the internal variables $\alpha$ to be discovered by means of this framework.}
    \label{fig:Overview}
\end{figure}

In this work, we propose a machine learning strategy to guide the discovery of internal variables and associated continuum models from atomistic/particle dynamics, here chosen to be described by Langevin dynamics. The proposed approach is based on two main considerations. First, the instantaneous value of the state variables shall be sufficient to predict the future state of the system, i.e., no memory effects. Second, the state variables ($\boldsymbol{\chi}$, $\boldsymbol{\alpha}$) shall fully describe the state of the system at any time, and hence we assume that the probability distribution of the atomistic/particle's positions is entirely determined by the current value of the state variables \cite{leadbetter2023statistical}. Toward these goals, we combine the information bottleneck (IB) method \cite{alemi2016deep,tishby2000information,wang2019past}, conditional normalizing flows (CNFs) \cite{huang2018neural,winkler2019learning}, and Variational Onsager Neural Networks (VONNs) \cite{huang2022variational} and denote this strategy as the IB-VONNs framework; see Fig.~\ref{fig:Overview}. The IB method is, in essence, an autoencoder with a loss function that takes into account the trade-off between the predictability and the complexity of the latent variables, both of which are quantified by means of the mutual information\footnote{This concept from information theory quantifies the amount of information that one variable contains about the other. It is intimately related to the notion of entropy and it enables the mathematical formulation of important thermodynamic concepts, such as Landauer's principle and the second law of thermodynamics, for processes that occur far away from equilibrium \cite{still2012thermodynamics,sagawa2019second}}. For our purposes, the input of the encoder is the present microscopic state of the system, and the goal is to reconstruct the microscopic state through the decoder, here using, in its more general version, CNFs. The IB and CNFs architectures are jointly trained with VONNs to learn the evolution equations so as to guarantee the Markovianity of the dynamics of the state variables as well as their thermodynamic consistency. We argue that the latent variables of the encoder-decoder architecture represent a good choice of internal variables, and that they enable the characterization of the probability distribution away from equilibrium. 

We note that the IB method (not combined with CNFs or VONNs) has already been successful in identifying reaction coordinates \cite{wang2019past} in biomolecular simulations, which are low-dimensional representations that are crucial to the pathways and configurations of molecules. This prior success and the analogy between reaction coordinates and internal variables suggest that the IB method could be a promising approach to discover internal variables in the context of mechanical applications. In this work, we modify the traditional encoder in the IB framework to guarantee that the internal variables (which are macroscopic in nature) are  permutation invariant with regard to various realizations, as is standard in statistical mechanics. We further combine it with CNFs, in order to further learn the microscopic probability distribution as a function of the enlarged state variables (this may significantly deviate from the Boltzmann distribution away from equilibrium). One key challenge in integrating these methods with machine learning approaches to learn PDEs lies in encoding the spatial dependency of the evolution equations on the state variables. For example, in Example 2 (Section \ref{sec: case2}), the evolution of the mean strain depends on local information and that of the nearest neighbors, while the evolution of internal variable only depends on local information. Such dependency is automatically encoded in the choice of the state and process variables through Onsager's variational principle, which lies at the heart of VONNs. 

We remark that, in theory, it would also be possible to first learn the internal variables and then leverage
a physics-based approach to derive the dynamics, such as Stochastic Thermodynamics with Internal Variables (STIV) \cite{leadbetter2023statistical}. Advances are however still required to computationally adapt STIV to general probability distributions of the kind learned via the framework proposed in this work (Section \ref{sec: IB-CNF}). We thus here use a data-driven approach to learn the evolution equations. In particular, we adopt VONNs \cite{huang2022variational} to learn the dynamics, as it strongly encodes thermodynamic constraints, it has strong statistical mechanics foundation and presents all the advantages of a completely variational approach (see \cite{huang2022variational} for further details). 


The proposed IB-VONNs method is applied in two examples. The first one is a toy problem: a single particle governed by the overdamped Langevin dynamics as a prototype for a colloidal particle in an optical trap. Although the problem is simple and the probability distribution is Gaussian at all times, it contains most of the key ingredients of a general mechanics problem and it is analytically tractable, thus making it a good first example on which to validate the proposed approach. Next, we apply the method to a one-dimensional system undergoing a phase transition, which may be considered as a prototype for a biological macromolecule or a nanorod, and whose macroscopic description still lacks a statistical mechanics foundation under general condition. In this example, the interparticle potential is multi-welled, and hence the density of states is multimodal. This density is first approximated with a Gaussian-mixture model, and then with CNFs to showcase the potential to model complex probability distributions for which we may lack physical intuition. The results indicate that the proposed framework is capable of automatically discovering internal variables that are physically based and lead to thermodynamically consistent evolution equations.

This paper is organized as follows. In Section \ref{sec: methods}, we discuss the desired properties that internal variables should have (Section \ref{Sec:InternalVariables}) and present all the key ingredients of the methodology developed in this work. These include the IB method and its variational formulation (Section \ref{sec: vib}), VONNs to learn the evolution equations (Section \ref{Sec:VONNs}), and CNFs to approximate arbitrary microscopic distributions (Section \ref{sec: CNF}). The proposed framework is presented in Section \ref{sec: IB-CNF}. In Sections \ref{sec: case1} and \ref{sec: case2}, the two aforementioned examples are discussed in detail, and the predicted probability distributions and macroscopic observables are compared to those obtained from multiple realizations of direct particle simulations. Finally, conclusions and future directions are discussed in Section \ref{sec: conclusions}.

\section{Methodology} \label{sec: methods}
\subsection{Internal variables} \label{Sec:InternalVariables}
The absence of a well-accepted statistical mechanics framework far away from equilibrium has naturally induced a lack of precise guidelines for the properties that internal variables should have in a thermodynamics formalism. Some broad considerations have however been summarized in the Roadmap on Multiscale Materials Modeling by~\cite{van2020roadmap}, where it was stated that macroscopic models should be (i) predictive, (ii) as simple as possible, (iii) formulated with physically intuitive variables, and they should (iv) establish links with lower scales to provide a mechanistic understanding and aid material design. 

We here refine these requirements so as to translate them into precise mathematical statements that can be incorporated into an automated framework for the discovery of internal variables. In particular, we require that the state variables at any given time (a) are capable of predicting the future behavior without requiring their past history
\cite{ottinger2005beyond,coleman1967thermodynamics,maugin1994thermodynamics}, 
(b) are as simple as possible, and (c) sufficient to characterize the salient features of the microscopic probability distribution at that time \cite{leadbetter2023statistical}. Furthermore, internal variables $\boldsymbol{\alpha}$ should be (d) 
 macroscopic quantities, measurable \cite{maugin1999thermomechanics,ottinger2005beyond} from the microscopic state $\boldsymbol{x}(t)$ and invariant with respect to the order of the realizations. It should thus be of the form 
\begin{equation} \label{Eq:InternalVariableEnsemble}
    \boldsymbol{\alpha} = \rho(\left< h(\boldsymbol{x}(t)) \right>),
\end{equation}
where $\left<\cdot\right>$ denotes the ensemble average, and $\rho$ and $h$ are some functions, potentially high-dimensional. 

We here use the information bottleneck method, described next, to automatically discover the internal variables.

\subsection{The information bottleneck method } \label{sec: vib}
The principle of IB was proposed by \cite{tishby2000information} as a variational principle for extracting the most relevant information $Z$ of a given signal $X$ that is most predictive of another signal $Y$($X$, $Y$ and $Z$ are, in general, random variables, and $x$,$y$,$z$ are instances of such variables, respectively). These signals can be understood in a general sense, and can refer to images, speech, particle trajectories or other data of interest. In the IB principle, the resulting optimization problem \cite{tishby2000information, still2014information} is written using the notion of mutual information $I$ between two random variables as 
\begin{equation} \label{Eq:BottleneckInference}
\max_{p(z|x)} \mathcal{L}[p(z|x)]=\max_{p(z|x)} \left[\underbrace{I[Z;Y]}_{\text{predictability}} - \gamma \underbrace{I[Z;X]}_{\text{complexity}} \right].
\end{equation}
The compromise between predictability and complexity is controlled by the parameter $\gamma \in [0,\infty)$, which is a user-defined non-negative constant. We recall that the mutual information $I[X;Y]$ between two variables quantifies the amount of information that one variable contains about the other through the joint and marginal probability distributions \cite{cover1999elements} as
\begin{equation}
    I(X;Y) = \iint p(x,y) \log \frac{p(x,y)}{p(x)p(y)} dx dy.
\end{equation}
Intuitively, to maximize \eqref{Eq:BottleneckInference}, we want to find a compressed representation, or latent variable, $Z$ of the original signal $X$ such that $Z$ preserves maximum information about the target signal $Y$, while being as simple as possible.

As a practical computational strategy, we adopt the Variational Information Bottleneck method (VIB) \cite{alemi2016deep}. The main idea is to use neural networks to model the encoder and decoder, and to approximate the mutual information, which is generally intractable, using the principle of variational inference. We assume that the latent variable $Z$ is a deterministic function of $X$, and, for simplicity, we drop the term $I(Z;X)$ in the loss function $\mathcal{L}$ and control the complexity of $Z$ by the neural network architecture. The optimal decoder $p(y|z)$ is approximated by $q(y|z)$ via a neural network. With this variational inference technique, the goal function can be approximated by a log-likelihood function, the proof of which can be found in \cite{alemi2016deep}; the mathematical form of the loss function will be given in detail in Section \ref{sec: IB-CNF} when the full IB-VONNs framework is presented.

\begin{figure}[ht]
    \centering
    \includegraphics[width=0.4\textwidth]{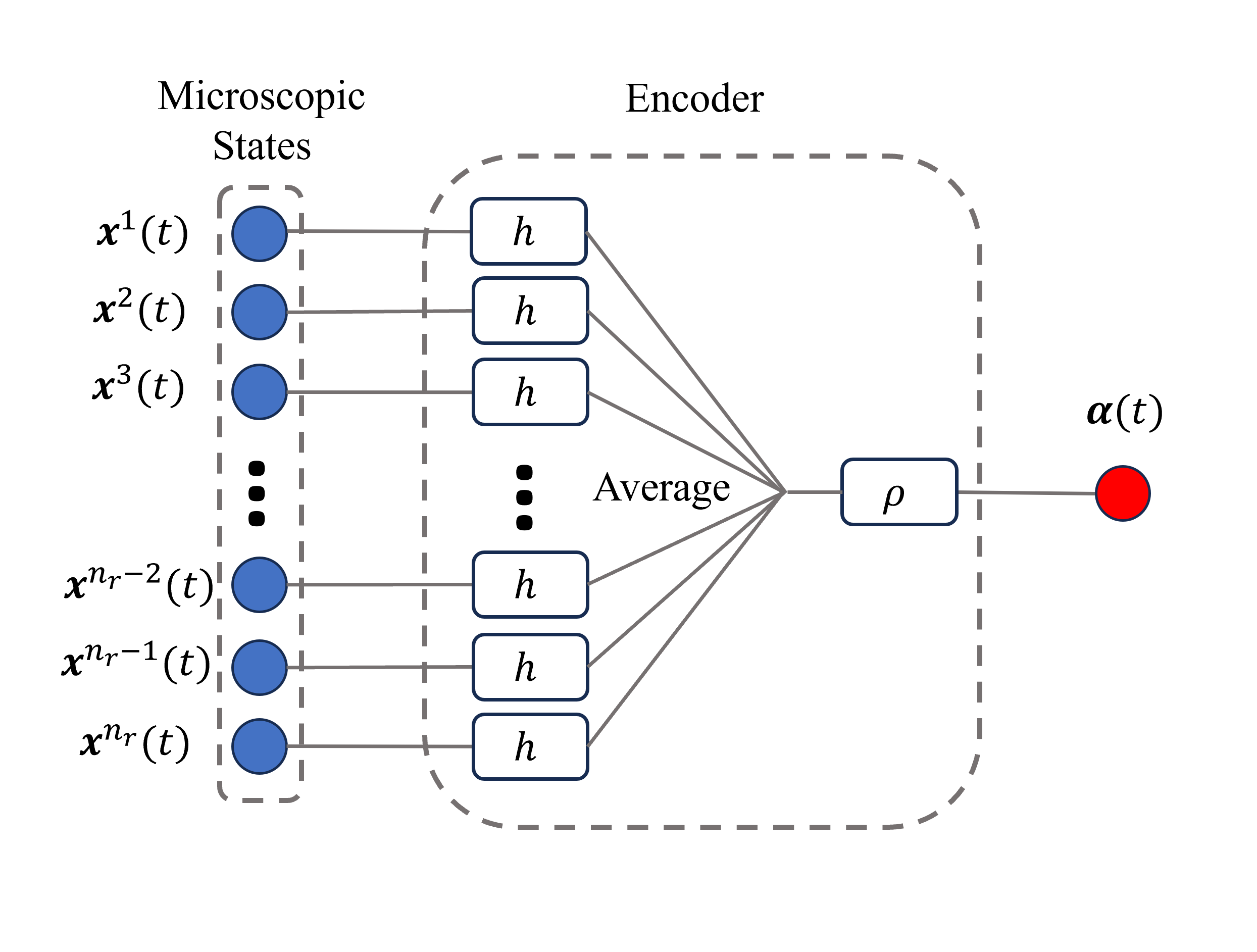}
    \caption{A schematic plot of the encoder architecture, used to learn the internal variables $\boldsymbol \alpha$. Such variables are expressed as $\boldsymbol{\alpha} = \rho \left(\frac{1}{n_r} \sum_{i=1}^{n_r} h(\boldsymbol{x}^i(t)) \right)$, with $h$ and $\rho$ modeled with neural networks, in order to ensure invariance with respect to the input order of the realizations. Here, $n_r$ denotes the number of realizations.}
    \label{fig: encoder}
\end{figure}

In this work, $X$ and $Y$ are chosen to be identical and represent the microscopic state of the system at some time $t$, $\boldsymbol{x}(t)$, and $Z$ encompasses the internal variables $\boldsymbol{\alpha}(t)$ determined by the encoder of the form of Eq.~\eqref{Eq:InternalVariableEnsemble} together with other known usual state variables $\boldsymbol{\chi}(t)$ (whose dependence on the microstate is known), i.e., $\boldsymbol{z}=(\boldsymbol{\chi}, \boldsymbol{\alpha})^T$. More specifically, the functions $h$ and $\rho$ are modeled by neural networks, and the ensemble average is numerically approximated from $N$ realizations as
\begin{equation}
    \boldsymbol{\alpha} = \rho \left(\frac{1}{n_r} \sum_{i=1}^{n_r} h(\boldsymbol{x}^i(t)) \right),
\end{equation}
where the superscript $i$ indicates the i-th realization and $n_r$ is the total number of realizations (see Fig. \ref{fig: encoder}). We would like to point out that this is a special case of the so-called Deep Sets architecture \cite{wagstaff2022universal,zaheer2017deep}. Since the order of different realizations contains no information, the encoder is learning functions on a set. It has been shown \cite{wagstaff2022universal,zaheer2017deep} that the above architecture is a universal approximator of set functions. Hence, the proposed encoder provides sufficient representability to discover internal variables that are direct functions of the microscopic states. 

For the decoder, $q(\boldsymbol{x}(t)|\boldsymbol{\alpha}(t), \boldsymbol{\chi}(t))$, which approximates the microscopic probability distribution, we first assume that it belongs to a certain family of distributions parameterized by a finite set of parameters, e.g., a Gaussian distribution in example 1 (Section \ref{sec: case1 learn iv}); and a Gaussian mixture model (GMM) in example 2 (Section \ref{Sec:case2_IB_VONNs}), see Fig.~\ref{fig: GMM}. Such an approach is simple and computationally efficient, though it is only of practical utility when there is intuition on the class of expected probability distributions throughout the system's evolution. In the interest of generality, we then introduce CNFs (described next), which allow learning arbitrary microscopic distributions from data, albeit at the expense of an increased computational cost. 

Although the internal variables discovered are approximated by neural networks and hence are not directly interpretable, we can gain insights on them by sampling microscopic states consistent with given values of the macroscopic state variables. This reconstruction network, i.e., the decoder, enables to go from the macroscopic picture to the microscopic one in a probabilistic sense, in full analogy to equilibrium statistical mechanics. This capability is naturally linked to the predictive capability of the state variables (including the internal variables) in the absence of memory effects. In particular, if the state variables were capable of exactly reproducing the microscopic probability distribution at all times, exact Markovian evolution equations for the state variables would exist \cite{leadbetter2023statistical}. However, it is important to remark that, an accurate representation of the the macroscopic description of the system does not always necessitate an accurate characterization of the microscopic probability distribution at all times. As an example of this, it was shown in \cite{leadbetter2023statistical} that a Gaussian approximation of the probability distribution lead to a highly accurate macroscopic description of a phase transforming system, whose true distribution is obviously multi-peaked. To ensure predictability as well as thermodynamic consistency, we here couple the learning of the internal variables, to that of the evolution equations. More specifically, the microscopic and macroscopic description at all times will be linked through the IB (encoder-decoder) architecture just described, and the value of the state variables $\boldsymbol{\alpha}, \boldsymbol{\chi}$ at consecutive time steps will be linked through the evolution equations, in this case, Variational Onsager Neural Networks (see Section \ref{Sec:VONNs}).

\subsubsection{Conditional normalizing flows} \label{sec: CNF}

In this section, we introduce CNFs \cite{huang2018neural, winkler2019learning} to tackle the task of learning arbitrary microscopic distribution $q(x(t)|\boldsymbol{z}(t))$, where $\boldsymbol{z}=(\boldsymbol{\chi}, \boldsymbol{\alpha})^T$. For simplicity, we only discuss the one-dimensional case, i.e., $x(t) \in \mathbb{R}^1$, though the architecture can be easily generalized to arbitrary dimensions. The idea of CNFs is to learn the target distribution by applying an invertible transformation on a simple distribution, here denoted as the base distribution $q(y|\boldsymbol{z}(t))$. We choose the base distribution to be a standard Gaussian distribution 
\begin{equation*}
    q(y|\boldsymbol{z}(t)) = \mathcal{N}(0,1).
\end{equation*}
Let $y = y(x(t), \boldsymbol{z}(t))$ be an invertible transformation with respect to $x(t)$. Then by the formula of change of variables,
\begin{equation*}
    q(x(t)|\boldsymbol{z}(t)) = q(y|\boldsymbol{z}(t)) \left| \det \left( \frac{\partial y}{\partial x(t)} \right) \right|,
\end{equation*}
where $\det \left( \frac{\partial y}{\partial x(t)} \right)$ is the determinant of the Jacobian of $y$ with respect to $x(t)$. 

\begin{figure}[ht]
     \centering
     \begin{subfigure}{0.49\textwidth}
         \centering
         \includegraphics[width=\textwidth]{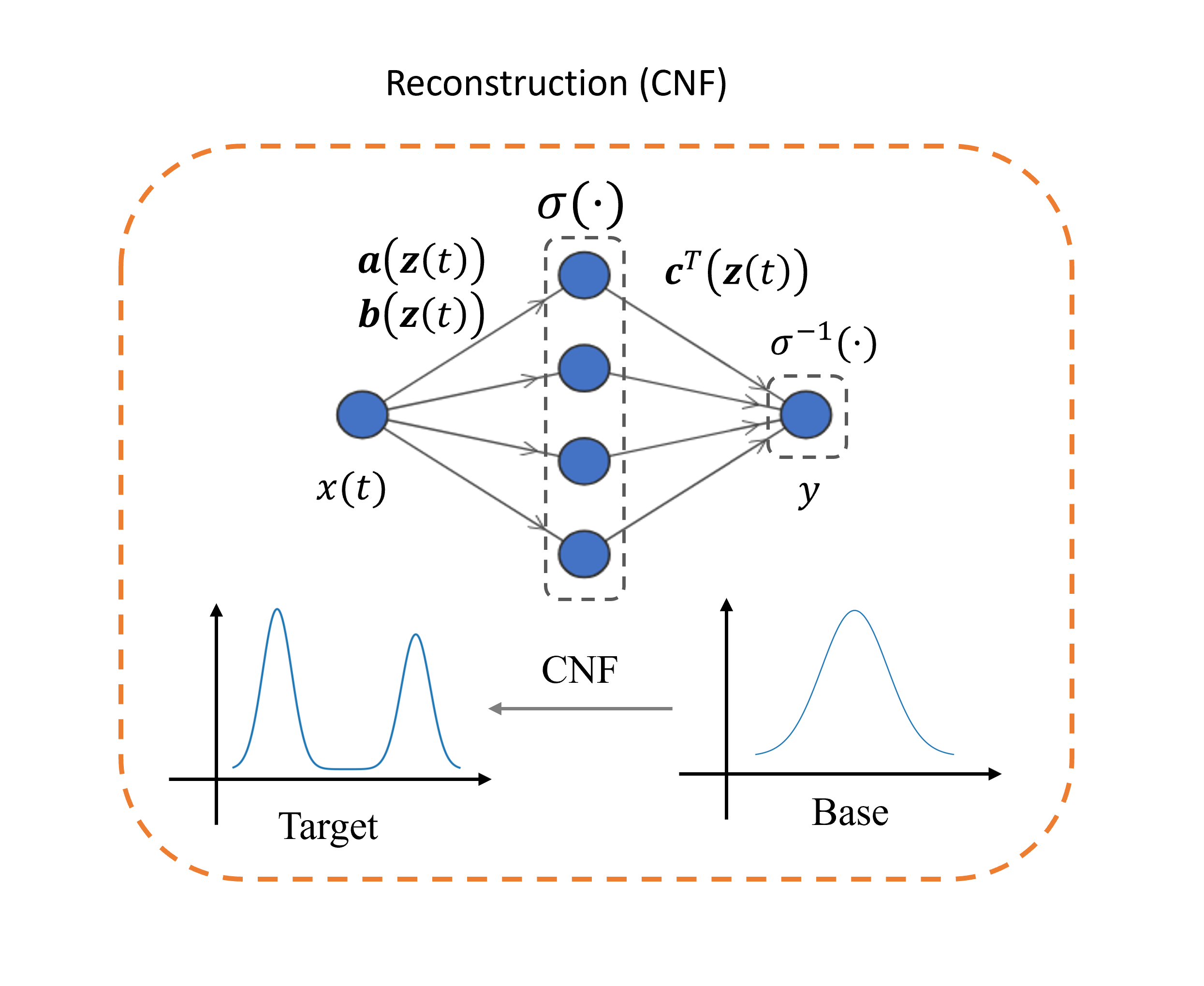}
         \caption{}
         \label{fig: CNF}
     \end{subfigure}
     \begin{subfigure}{0.49\textwidth}
         \centering
         \includegraphics[width=\textwidth]{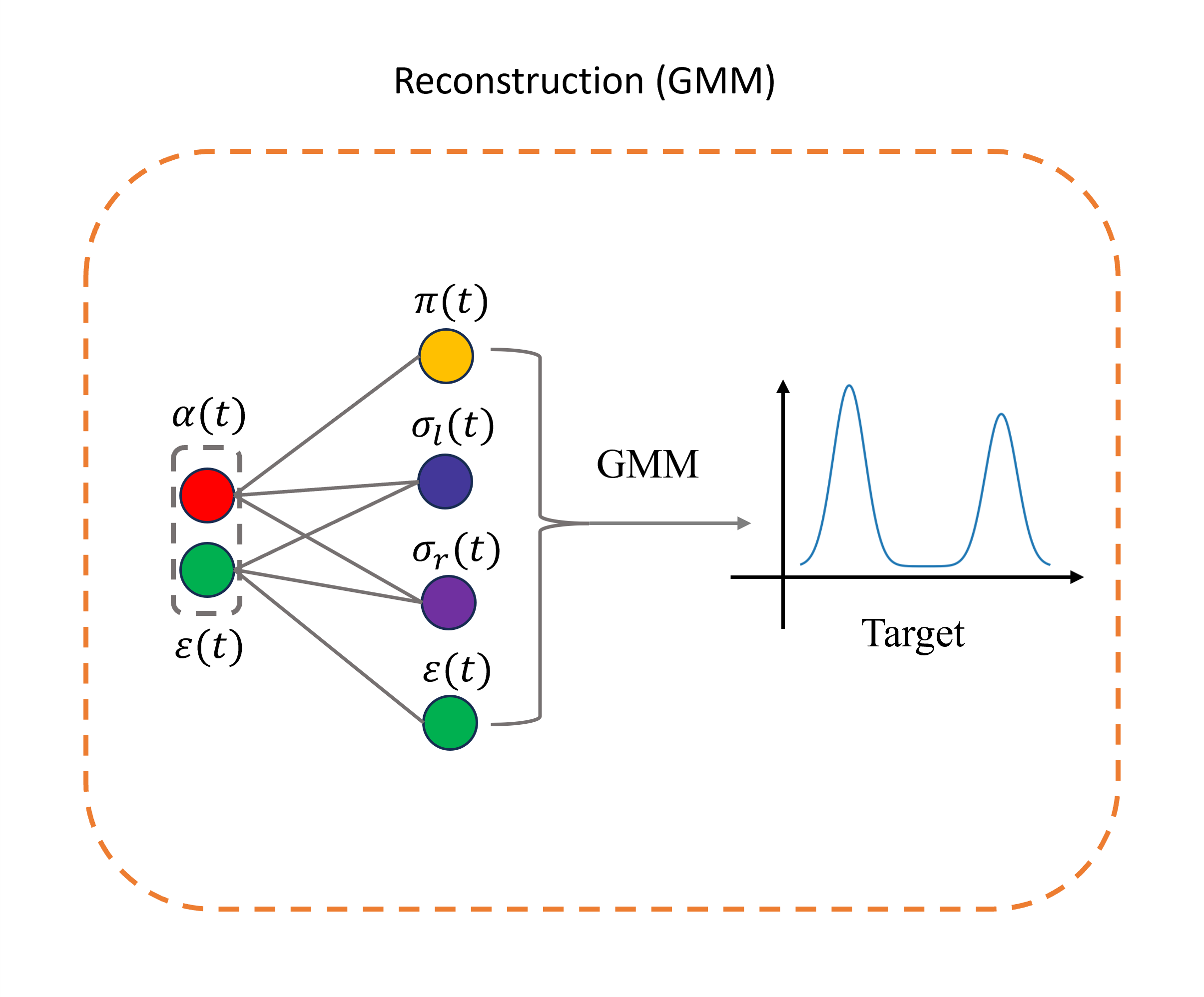}
         \caption{}
         \label{fig: GMM}
     \end{subfigure}
\caption{(a) A schematic plot of the CNFs decoder architecture (with $d=4$). The parameters $\boldsymbol{a}$, $\boldsymbol{b}$ and $\boldsymbol{c}$ are functions of the state variables $\boldsymbol{z}(t)$, and they are modeled by neural networks. $\sigma(\cdot)$ and $\sigma^{-1}(\cdot)$ are the sigmoid and logit activation functions, respectively. (b) A schematic plot of the GMM decoder architecture used in Example 2. It is composed of two Gaussians and parameterized by the mean strain $\varepsilon$, the variances of the two Gaussians, $\sigma_l$ and $\sigma_r$, and the mixing coefficient $\pi$.}
\label{fig: CNF_GMM}
\end{figure}

The key point is to parameterize the invertible mapping $y$. For this, we adopt the deep sigmoidal flows architecture proposed in \cite{huang2018neural}, which has been shown to be a universal approximator to any invertible transformation. The mapping $y$ is modeled by a neural network-like structure with one hidden layer, see Fig.~\ref{fig: CNF},
\begin{equation*}
    y = y(x(t), \boldsymbol{z}(t)) = \sigma^{-1} \left(\boldsymbol{c}^T \sigma(\boldsymbol{a} x(t) + \boldsymbol{b}) \right),
\end{equation*}
where the parameters $\boldsymbol{c}=\boldsymbol{c}(\boldsymbol{z}(t)), \boldsymbol{a}=\boldsymbol{a}(\boldsymbol{z}(t))$, and $\boldsymbol{b}=\boldsymbol{b}(\boldsymbol{z}(t)) \in \mathbb{R}^{d\times 1}$ are functions of the state variables and modeled by neural networks, $\sigma(\cdot)$ is the element-wise sigmoid activation function, and $\sigma^{-1}(\cdot)$ is the inverse of the sigmoid, or the logit function. We require the elements of $\boldsymbol{c}$ to be between 0 and 1, and sum up to 1, and that the elements of $\boldsymbol{a}$ are strictly positive to make sure $y$ is invertible \cite{huang2018neural}. $d$ is the number of hidden units chosen by the user (in Fig.~\ref{fig: CNF}, $d=4$). With this architecture, the approximated target distribution can be explicitly written as
\begin{align*}
    q(x(t)|\boldsymbol{z}(t)) = p(y|\boldsymbol{z}(t)) \cdot (\sigma^{-1})^{'} \left(\boldsymbol{c}^T \sigma(\boldsymbol{a} x(t) + \boldsymbol{b}) \right) \cdot \boldsymbol{c}^T \left(\boldsymbol{a} \odot \sigma^{'}(\boldsymbol{a} x(t) + \boldsymbol{b})\right),
\end{align*}
where "$\odot$" is the element-wise product, or the Hadamard product. The loss function is chosen to be the negative log-likelihood; see Section \ref{sec: IB-CNF}.

\subsection{Variational Onsager Neural Networks} \label{Sec:VONNs}
VONNs \cite{huang2022variational} is a variational strategy for learning thermodynamically consistent evolution equations, by directly learning the action density guiding the evolution. It is based on Onsager's variational principle \cite{doi2011onsager}, which for isothermal systems with negligible inertia reads as 
\begin{equation} \label{eq: OVP}
    \frac{\delta \mathcal{R}}{\delta \boldsymbol{w}} = \boldsymbol{0},
\end{equation} 
where $\boldsymbol{z}$ are the state variables, i.e. $\boldsymbol{z}=(\boldsymbol{\chi}, \boldsymbol{\alpha})^T$, $\boldsymbol{w}$ are the process variables describing how the system dissipates energy, and $\mathcal{R}$ is the Rayleighian, defined as 
\begin{equation}
    \mathcal{R}(\boldsymbol{z},\boldsymbol{w}) = \dot{\mathcal{F}}(\boldsymbol{z}) + \mathcal{D}(\boldsymbol{z},\boldsymbol{w}) + \mathcal{P}(\boldsymbol{z},\boldsymbol{w}).
\end{equation}
Here, $\mathcal{F}$ is the free energy, $\mathcal{D}$ is the dissipation potential, and $\mathcal{P}$ is the power of the external forces. As in \cite{huang2022variational}, we assume the existence of densities $f$ and $\psi$ for the free energy and dissipation potentials, i.e., $\mathcal{F} = \int f\, dV, \, \mathcal{D} = \int \psi\, dV$, which satisfy the following conditions for thermodynamic consistency: (i) $\psi(\boldsymbol{z},\boldsymbol{w})$ convex in $\boldsymbol{w}$; (ii) $\psi(\boldsymbol{z},\boldsymbol{0}) = 0$; (iii) $\frac{\partial \psi}{\partial \boldsymbol{w}}\Big|_{\boldsymbol{w}=\boldsymbol{0}} = \boldsymbol{0}$. Under these conditions, and in the absence of external power $\mathcal{P}$, the rate of change in the free energy is non-positive, in accordance with the second law of thermodynamics.

In practice, solving the evolution equation derived form Eq.~\eqref{eq: OVP} usually involves computing the inverse of $\frac{\partial \psi}{\partial \boldsymbol{w}}$, which is computationally expensive and inconvenient. In this work, we consider a variant of the original formulation to avoid computing such inverse. We define the dual dissipation potential density $\phi$ to be the Legendre transform of the dissipation potential density 
\begin{equation}
    \phi(\boldsymbol{z},\boldsymbol{g}) = \boldsymbol{w}(\boldsymbol{z},  \boldsymbol{g}) \cdot \boldsymbol{g} - \psi\left(\boldsymbol{z}, \boldsymbol{w}( \boldsymbol{z}, \boldsymbol{g}) \right),
\end{equation}
where $\boldsymbol{g} = \frac{\partial \psi}{\partial \boldsymbol{w}}$ is invertible with respect to $\boldsymbol{w}$, since $\psi(\boldsymbol{z},\boldsymbol{w})$ is convex on $\boldsymbol{w}$, and $\boldsymbol{w}({\boldsymbol{z},} \boldsymbol{g})$ is the inverse function of $\boldsymbol{g}$ with respective to $\boldsymbol{w}$, satisfying
\begin{equation}
    \boldsymbol{w} = \frac{\partial \phi}{\partial \boldsymbol{g}}.
\end{equation}
Correspondingly, the thermodynamic constraints on $\phi$ read: (i) $\phi(\boldsymbol{z},\boldsymbol{g})$ is convex on $\boldsymbol{g}$; (ii) $\phi(\boldsymbol{z},\boldsymbol{0}) = 0$; (iii) $\frac{\partial \phi}{\partial \boldsymbol{g}}\Big|_{\boldsymbol{g}=\boldsymbol{0}} = \boldsymbol{0}$. Equation \eqref{eq: OVP} is a set of differential equations of the form \cite{huang2022variational}
\begin{equation}
    \frac{\partial \psi}{\partial \boldsymbol{w}} = \mathcal{N}_{\boldsymbol{z}}(f),
\end{equation}
where $\mathcal{N}_{\boldsymbol{z}}$ is an operator, and $\mathcal{N}_{\boldsymbol{z}}(f)$ is a function of the state variables. It then follows that the process variables can be directly computed as a function of the state variables as
\begin{equation} \label{Eq:VONNs_w_eq}
    \boldsymbol{w} = \frac{\partial \phi}{\partial \boldsymbol{g}}\left(\boldsymbol{z}, \mathcal{N}_{\boldsymbol{z}}(f) \right),
\end{equation}
from $f$ and $\phi$ without needing to compute any inverse function. Usually, the derivative of the state variables with respect to time, $\dot{\boldsymbol{z}}$, can be computed from $\boldsymbol{w}$ with little effort. Then, we can integrate $\dot{\boldsymbol{z}}$ to obtain the trajectories of the state variables, and define the loss function as the $L^2$ error between the predicted trajectories and the ground truth; see Section \ref{sec: IB-CNF} for more details. In this way, the inverse function of $\boldsymbol{g}$ does not need to be computed, neither during the training nor for predicting the dynamics. 

In VONNs, an Integrable Neural Network (INN) \cite{teichert2019machine,teichert2020scale} and a Partially Input Convex Integrable Neural Network (PICINN) \cite{amos2017input, huang2022variational} are used to represent the non-dimensional free energy density $\tilde{f}$ and the non-dimensional dual dissipation potential density $\tilde{\phi}$, respectively. The densities $f$ and $\phi$ are then recovered as:
\begin{align} \label{Eq:f_VONNs}
    &f(\boldsymbol{z}) = f^{*} (\tilde{f}(\boldsymbol{z}) - \tilde{f}( \boldsymbol{0})), \\ \label{Eq:psi_VONNs}
    &\phi(\boldsymbol{z},\boldsymbol{g}) = \phi^{*} \left( \tilde{\phi}(\boldsymbol{z},\boldsymbol{g}) - \tilde{\phi}(\boldsymbol{z},\boldsymbol{0}) - \frac{\partial \tilde{\phi}}{\partial \boldsymbol{g}}\Big|_{\boldsymbol{g}=\boldsymbol{0}} \cdot \boldsymbol{g} \right),
\end{align}
where  $f^{*}$ and $\phi^{*}$ are characteristic scales. Equation \eqref{Eq:psi_VONNs} together with the PICINN architecture ensures that the three aforementioned conditions on $\phi$ are strongly encoded, thus guaranteeing compliance with the second law of thermodynamics.

\subsection{IB-VONNs} \label{sec: IB-CNF}
\begin{figure}[ht]
    \centering
    \includegraphics[width=0.9\textwidth]{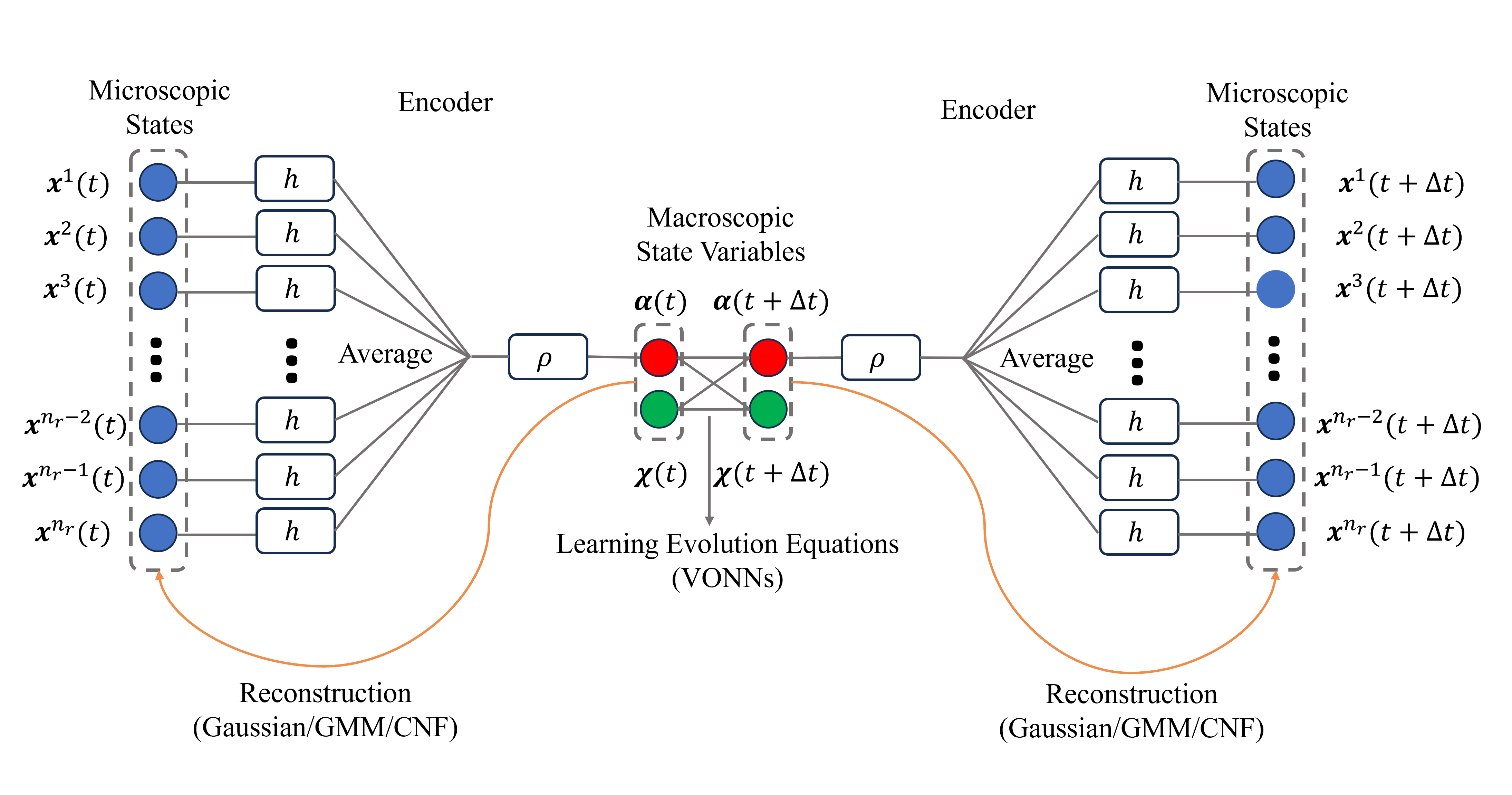}
    \caption{A schematic plot of the IB-VONNs architecture. It can be understood as an IB structure to learn the internal variables of the system, augmented with VONNs to learn the evolution equations of the state variables $\boldsymbol{\alpha}$ and $\boldsymbol{\chi}$. The two autoencoders in the figure refer to one single encoder and decoder applied on data at different times. }
    \label{fig:IB-CNF}
\end{figure}

In this section, we describe the architecture of the proposed framework, which combines the IB method to discover the internal variables and VONNs to ensure the Markovianity of the state variables as well as the thermodynamic consistency of the learned evolution equations. We denote this as the IB-VONNs framework (Fig. \ref{fig:IB-CNF}). 

In this framework, the first sub-task consists of inferring the microscopic probability distribution function (pdf) at any simulated time $s$ from the macroscopic state variables at that time, i.e., learning the conditional distribution $q(x(s)|\boldsymbol{z}(s))$ ($s \in [t_0,t_f]$, where $t_0$ and $t_f$ are the initial and final time, respectively). We first assume this distribution belongs to certain family, namely, a Gaussian in Example 1 (Section \ref{sec: case1 learn iv}) and a GMM in Example 2 (Section \ref{Sec:case2_IB_VONNs}), and then we apply CNFs (Section \ref{sec: case 2 IB_CNF}), which is a universal approximator and does not require prior knowledge of the microscopic physics. We use the mutual information loss (approximated by the log-likelihood) as the reconstruction loss function
\begin{equation} \label{Eq:IB_loss}
    \mathcal{L}_{pdf} = -\frac{1}{N_{pdf}} \sum_{i=1}^n \sum_{j=0}^{n_t} \sum_{k=1}^{n_r} \log q(x_i^k(t_0 + j\cdot \Delta t)|\boldsymbol{z}_i(t_0 + j\cdot \Delta t)),
\end{equation}
where indices $i\in\{1,2,\ldots,n\}$, $j\in\{1,2,\ldots,n_t\}$ and $k\in\{1,2,\ldots,n_r\}$ are integers used to label the spatial and temporal collocation points and the realizations of the stochastic microscopic dynamics, respectively, $n_t \cdot \Delta t = t_f-t_0$, and $N_{pdf}=n(n_t+1)n_r$ is the number of total data points in the time interval $t_f-t_0$. 

The second sub-task is aimed at updating the macroscopic variables $\boldsymbol{z}(t)=(\boldsymbol{\chi}(t),\boldsymbol{\alpha}(t))$ with VONNs. 
In the examples considered the particles obey Langevin dynamic equations, and hence, the inputs to VONNs will be intrinsically noisy induced by the necessarily finite number of realizations (this is contrast to the cases studied in the original paper of VONNs \cite{huang2022variational}, where all examples used data from deterministic simulations). To increase the robustness of VONNs in this scenario, we adopt an integral formulation, where deviations in the predictions are quantified with the following $L^2$ loss functions 
\begin{equation} \label{Eq:loss_chi}
    \mathcal{L}_{\boldsymbol{\chi}} = \frac{1}{N} \sum_{i=1}^n \sum_{j=1}^{n_t} \big| \boldsymbol{\chi}^{NN}_i(t_0 + j\cdot \Delta t) - \boldsymbol{\chi}_i(t_0 + j \cdot \Delta t) \big|^2, 
\end{equation}
\begin{equation} \label{Eq:loss_alpha}
    \mathcal{L}_{\boldsymbol{\alpha}} = \frac{1}{N} \sum_{i=1}^n \sum_{j=1}^{n_t} \big| \boldsymbol{\alpha}^{NN}_i(t_0 + j\cdot \Delta t) - \boldsymbol{\alpha}_i(t_0 + j \cdot \Delta t) \big|^2,
\end{equation}
where here and in the following we use the superscript $NN$ to indicate neural network prediction and $N=nn_t$ is the number of total data points. Further loss terms may be considered in the presence of other macroscopic observables (e.g., external forces), as will be done in the first example (see Section \ref{sec: case1}).

The three loss functions introduced above, Eqs.~\eqref{Eq:IB_loss}-\eqref{Eq:loss_alpha}, are combined to deliver the following loss function
\begin{equation}
    \mathcal{L} = \underbrace{\lambda_{pdf} \mathcal{L}_{pdf}}_{\text{IB}} + \underbrace{\lambda_{\boldsymbol{\chi}} \mathcal{L}_{\boldsymbol{\chi}} + \lambda_{\boldsymbol{\alpha}} \mathcal{L}_{\boldsymbol{\alpha}}}_{\text{VONNs}},
\end{equation}
which will be used to simultaneously train all the neural networks. We note that the proper choice of the loss weights,  $\lambda_{pdf}, \lambda_{\boldsymbol{\chi}}, \lambda_{\boldsymbol{\alpha}}$, is crucial to the success of training. However, to the best of our knowledge, none of the existing methods to determine the loss weights automatically 
\cite{mcclenny2023self,wang2022and} work sufficiently well for the problems here studied, and, hence, these are hand-tuned in the following examples. Further investigation on this topic is beyond the scope of this work, but it is of practical and theoretical interest. We point out that reconstructing the marginal distribution in the two examples considered in this paper is not necessary, and one could, in theory, set $\lambda_{pdf}$ to be zero without affecting the capability of the IB-VONNs framework to discover internal variables. Nevertheless, the reconstruction part is useful to interpret the meaning of the discovered internal variables. 

We remark that the IB-VONNs method naturally encodes the requirements set for internal variables discussed in Section \ref{Sec:InternalVariables}: (a) the state variables are chosen to be maximally predictive of the future state in the absence of memory effects (note that the input of VONNs only includes the present state); (b) the complexity of $\boldsymbol{\alpha}$ can be controlled with the architecture of the encoder and its dimensionality; (c) the reconstruction part ensures that the microscopic probability distribution is characterized by the state variables and (d) $\boldsymbol{\alpha}$ is computed from the microscopic state, following Eq.~\eqref{Eq:InternalVariableEnsemble}, hence ensuring that they are macroscopic quantities invariant with respect to the order of the realizations. 

\section{Example 1: a single particle system} \label{sec: case1}
\subsection{Model setup and simulation details} \label{sec: case1 setup}
\begin{figure}[ht]
    \centering
    \includegraphics[width=0.35\textwidth]{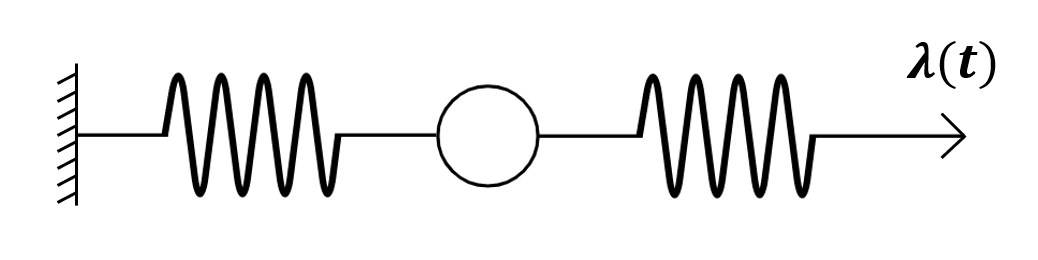}
    \caption{A prototypical model for a colloidal particle in an optical trap. The particle is connected to a fix point via a spring and pulled via a spring according to a pulling protocol $\lambda(t)$.}
    \label{fig:case1 particle}
\end{figure}

As an illustrative example of the method proposed in this work, we consider the prototypical model of a colloidal particle in an optical trap. This consists of a single particle, whose position $x(t)$ is governed by overdamped Langevin dynamics in a quadratic potential, see Fig.~\ref{fig:case1 particle},
\begin{equation} \label{eq:single par lagevin} 
    \eta \dot{x}(t)  = k (\lambda(t) - 2x(t)) + \sqrt{2\eta k_B T} \dot{\xi}.
\end{equation}
Here, $k$ is the spring constant, $\eta$ is the viscosity, and $\lambda(t)$ is an external pulling protocol. The particle is subjected to Brownian forces (last term in the equation), where $k_B$ is the Boltzmann constant, $T$ the temperature, and $\xi$ is a standard Brownian motion satisfying $\langle\dot{\xi}(t)\rangle=0$ and $\langle\dot{\xi}(t)\dot{\xi}(t^{'})\rangle=\delta(t-t^{'})$. This example contains all of the ingredients of a mechanically-relevant system, i.e., it is deformable and can be externally actuated, while being analytically solvable. It is thus ideal to test the proposed approach.

The probability distribution function $p(x,t)$ of the particle displacement satisfies the Fokker-Planck equation
\begin{equation}
    \frac{\partial p(x,t)}{\partial t} = -\frac{k}{\eta} \frac{\partial [(\lambda(t)-2x) p(x,t)]}{\partial x} + \frac{k_B T}{\eta} \frac{\partial^2 p(x,t)}{\partial x^2}, 
\end{equation}
which admits a Gaussian solution 
\begin{equation}
    p(x,t) = \frac{1}{\sqrt{2 \pi \sigma^2(t)}} \exp \left( - \frac{\left(x - \mu(t)\right)^2}{2 \sigma(t)^2} \right),
\end{equation}
whose mean $\mu(t)$ and variance $\sigma(t)$ evolve according to 
\begin{align} 
    \label{Eq:Case1_mean} 
    &\dot{\mu}(t) + \frac{(2\mu(t) - \lambda(t))k}{\eta} = 0  \\ 
    \label{Eq:Case1_variance}
    &\dot{\sigma}(t) + \frac{2k\sigma(t)}{\eta} - \frac{k_B T}{\eta \sigma(t)} = 0.
\end{align}
These equations are decoupled, and for initial conditions given by the equilibrium distribution, it can be analytically shown that the variance remains constant regardless of the pulling protocol (note that Eq.~\eqref{Eq:Case1_variance} does not depend on $\lambda)$. 
 
Each realization $x(t)$ of the above stochastic dynamics is considered microscopic, while a macroscopic observable will be some ensemble average $\left<\cdot\right>$ of the system. An obvious choice of (macroscopic) state variables is the mean $\mu(t)=\langle x(t)\rangle$ and the variance $\sigma(t)=\langle \left(x(t)-\mu\right)^2\rangle $. These can exactly capture the microscopic probability distribution, and hence their evolution equation is guaranteed to be Markovian \cite{leadbetter2023statistical}. 

In the following, we consider the special case in which the system is initialized at equilibrium (and hence the variance remains constant throughout the evolution), for which a single variable suffices to describe the macroscopic behavior of the system. We apply IB-VONNs with Gaussian assumption on the particle's trajectories, and check if it can discover a suitable state variable, i.e., some one-to-one mapping of the mean particle's displacement. 

We simulate $n_r=4000$ trajectories of the governing stochastic equation, Eq.~\eqref{eq:single par lagevin}, with parameters
\begin{equation*}
    k = 1, \quad \eta = 1, \quad k_B T = 10^{-2},
\end{equation*}
under the pulling protocol $\lambda(t) = vt(1 - e^{-t})(1 + 2 \sin(2 \pi t))$ for $t \in [0,3]$. The equation is solved with a uniform time step $\Delta t=0.01$ using the Euler-Maruyama scheme and initial conditions drawn from the equilibrium distribution. 
We consider 21 equally spaced pulling velocities $v_i$ ranging from 0.5 to 1.0, i.e., $v_i = 0.5 \cdot (1 + \frac{i}{20}), \, i=0,1,\cdots,20$. Data generated with pulling velocities $v_0, v_2, v_4 \cdots, v_{20}$ is used as the training set, while the rest forms the test set. 

\subsection{Macroscopic description} \label{Sec:Colloidal_MacroDescription}
For the purposes of this example, we will consider the pulling protocol (total displacement) $\lambda$ to be the macroscopic state variable describing the system at equilibrium. Outside of equilibrium, the state variables will then be $\boldsymbol{z}=(\lambda, \alpha)^T$, where the learned internal variable $\alpha$ is expected to be a one-to-one mapping to the mean position, i.e., $\alpha = \alpha(\mu)$, as just discussed. The rate of change of the internal variable $\dot{\alpha}$ is chosen to be the process variable. We then derive the evolution equations for the state variables by Onsager's variational principle (considering traction boundary conditions for such purpose). The Rayleighian reads
\begin{equation}
    \mathcal{R} = \dot{f} + \psi - F_{ex} \dot{\lambda},
\end{equation}
where $F_{ex}(t)=\left<k(\lambda(t) - x(t))\right>=k(\lambda - \mu)$ is the ensemble average of the external force exerted on the system, and $f=f(\lambda,\alpha)$ and $\psi=\psi(\alpha,\dot{\alpha})$ are the free energy and dissipation potential, respectively. By performing variation on the process variables $\dot{\lambda}$ and $\dot{\alpha}$, we find that the system obeys the following partial differential equations 
\begin{align} 
&\frac{\partial f}{\partial \lambda} = F_{ex}, \label{eq: case1 OVP fex}\\ 
&\frac{\partial f}{\partial \alpha} + \frac{\partial \psi}{\partial \dot{\alpha}} = 0  \label{eq: case1 OVP iv} .
\end{align}
Fixing the zero of the free energy as  $f(0,\alpha(\mu=0))=0$, and requiring, for thermodynamic consistency (see Section \ref{Sec:VONNs}), that $\psi(\alpha,0)=0$ and $\frac{\partial \psi}{\partial \dot{\alpha}}\Big|_{\dot{\alpha}=0} = 0$, we can find a unique analytic solution for the free energy and dissipation potential (see Appendix \ref{appx: case 1} for the proof of uniqueness)
\begin{equation} \label{eq: case1 fe}
    f = \frac{1}{2} k (\lambda - \mu)^2 + \frac{1}{2} k \mu^2, 
\end{equation}
\begin{equation}
    \psi = \frac{1}{2} \eta \dot{\mu}^2 = \frac{1}{2} \eta \left[ (\alpha^{-1})^{'}\dot{\alpha} \right]^2.
\end{equation}
Here, $\mu$ is expected to be a bijection of $\alpha$, i.e., $\mu = \alpha^{-1}(\alpha)$, where $\alpha^{-1}$ is the inverse of $\alpha(\mu)$, and $(\alpha^{-1})^{'}=\frac{d \alpha^{-1}}{d \alpha}$. The resulting evolution equation of $\alpha$ is consistent with that in Eq.~\eqref{Eq:Case1_mean}, namely,
\begin{equation} \label{eq: case1 evo alpha}
    \dot{\alpha} = \alpha' \dot{\mu} = - \frac{k}{\eta} \alpha' \cdot \left(2 \alpha^{-1}(\alpha) - \lambda\right),
\end{equation}
where $\alpha'$ is short for $\frac{d \alpha(\mu)}{d \mu}\Big|_{\mu=\left(\alpha^{-1}(\alpha)\right)}$.

In practice, as discussed in Section \ref{Sec:VONNs}, we will directly learn the dual dissipation potential $\phi(\alpha,g)$, which is the Legendre transform of $\psi(\alpha,\dot{\alpha})$ with respect to $\dot{\alpha}$, and hence $g=\frac{\partial \psi}{\partial \dot{\alpha}}$. 
Since it is not directly observable, we use Eq.~\eqref{eq: case1 OVP iv} to compute $g$ from data as $g \coloneqq \frac{\partial \psi}{\partial \dot{\alpha}} = -\frac{\partial f}{\partial \alpha}$, bypassing in this way the need to model $\psi$ with a neural network. Equation \eqref{eq: case1 OVP iv} can then be rewritten as (see Eq.~\eqref{Eq:VONNs_w_eq})
\begin{equation}
    \dot{\alpha} = \frac{\partial \phi(\alpha, g)}{\partial g}\Big|_{g=-\frac{\partial f}{\partial \alpha}},
\end{equation}
which may be integrated in time to obtain the evolution of the internal variable
\begin{equation}
    \alpha^{NN}(t) = \int_0^t \dot{\alpha}^{NN}(s) ds.
\end{equation}

\subsection{Learning the internal variable and evolution equations} \label{sec: case1 learn iv}

We use the IB-VONNs architecture of Fig.~\ref{fig:IB-CNF} with one internal variable and $\chi=\lambda$. The encoder $h$ is represented by a neural network with two hidden layers and $10$ neurons per layer, for which we use the hyperbolic tangent activation function, and the nonlinear transformation $\rho$ is modeled by a neural network with $1$ hidden layer with $10$ neurons. For simplicity, and also based on the prior knowledge of the system, we model the probability distribution $q(x(t) | \boldsymbol{z}(t))$ by a Gaussian distribution with mean $\mu^{NN}(t)$ and standard deviation $\sigma^{NN}(t)$. Such mean and standard deviation are computed as the outputs $\tilde{\mu}^{NN}$ and $\tilde{\sigma}^{NN}$ of a neural network with two hidden layers and $10$ neurons per layer, multiplied by the corresponding characteristic scales, $\mu^{*}$ and $\sigma^{*} $, i.e., 
\begin{equation}
    \mu^{NN}(t) = \mu^{*} \tilde{\mu}^{NN}(t) , \quad \sigma^{NN}(t) = \sigma^{*} \tilde{\sigma}^{NN}(t).
\end{equation}
Here, $\mu^{*}$ and $\sigma^{*}$ are chosen as the mean and the standard deviation of the particle's displacement in the training set. The non-dimensional free energy is represented by an INN with $2$ hidden layers with $20$ neurons each, and the non-dimensional dual dissipation potential is approximated by a PICINN with $2$ hidden layers, each with $20$ neurons per layer for the convex and non-convex portions of the neural network (i.e., $40$ neurons in total per layer). The characteristic scales of the free energy and dual dissipation potential densities in VONNs are estimated as (Eq.~\eqref{eq: case1 OVP fex} and \eqref{eq: case1 OVP iv})
\begin{equation}
    f^{*} = \sigma_{F_{ex}} \sigma_{\lambda}, \quad \phi^{*} = f^{*}.
\end{equation}
where $\sigma_{F_{ex}}$ and $\sigma_{\lambda}$ are the standard deviations of $F_{ex} $ and $ \lambda$, respectively (these are computed from data at all times from all pulling protocols, hence $\sigma_\lambda\neq 0$). These networks are trained with the following loss function
\begin{align}
    \mathcal{L} = &- \frac{\lambda_{pdf}}{N_{pdf}} \sum_{j=0}^{n_t} \sum_{k=1}^{n_r}  \log q(x^k(t_0 + j\cdot \Delta t) | \alpha(t_0 + j\cdot \Delta t))\\
    &+ \frac{\lambda_{F_{ex}}}{N} \sum_{j=1}^{n_t} \Big| \frac{\partial f\left(\lambda(t_0 + j\cdot \Delta t), \alpha(t_0 + j\cdot \Delta t)\right)}{\partial \lambda} - F_{ex}(t_0 + j\cdot \Delta t) \Big|^2 \\
    &+ \frac{\lambda_{\alpha}}{N} \sum_{j=1}^{n_t} \Big| \alpha^{NN}(t_0 + j\cdot \Delta t) - \alpha(t_0 + j\cdot \Delta t) \Big|^2
\end{align}
using an Adam optimizer with learning rate $5 \times 10^{-5}$. Here, $N=n_t$ and $N_{pdf}=(n_t+1)n_r$ are the number of training data points for the various loss terms, and $\lambda_{F_{ex}}=\lambda_{\alpha}=1$ and $\lambda_{pdf}=10^{-5}$ are the manually-chosen loss weights. Mini-batch training is adopted to reduce memory consumption and to avoid local minima. A batch consists of data within the time range of $[t_0, t_f]$, where the starting time $t_0$ is chosen randomly from $[0,3-(t_f-t_0)]$. $t_f-t_0$ is set to be $0.3$ initially, and increased to $3$ during the training process.

\begin{figure}[ht]
     \centering
     \begin{subfigure}{0.52\textwidth}
         \centering
         \includegraphics[width=\textwidth]{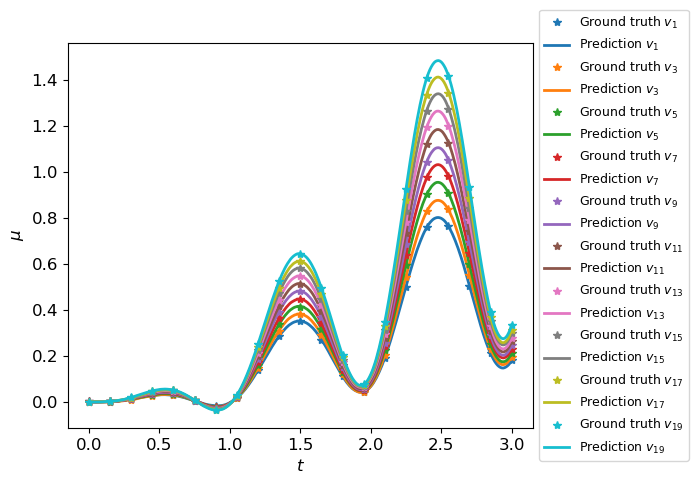}
         \caption{}
         \label{fig:case1 IB mu sigma}
     \end{subfigure}
     \begin{subfigure}{0.43\textwidth}
         \centering
         \includegraphics[width=\textwidth]{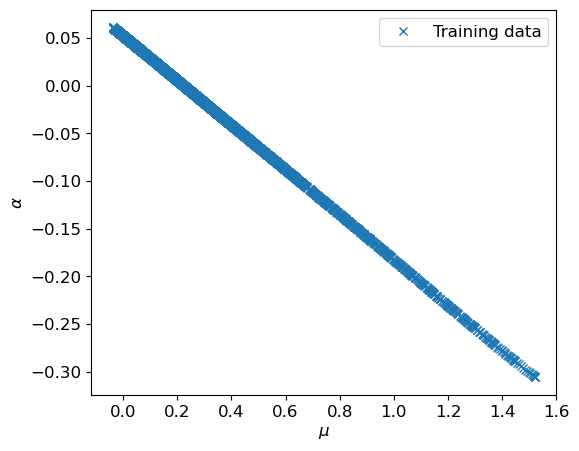}
         \caption{}
         \label{fig: case1 internal vs mean}
     \end{subfigure}
\caption{(a) Decoder predictions of the mean particle's displacement versus the ground truth (directly computed from the Langevin simulations) for the pulling velocities $v_i$ considered in the test set. (b) Learned internal variable versus the true mean displacement for the training data. The results perfectly collapse onto one single curve for all pulling velocities $v_i$.}
\label{fig: case1 sinusoidal}
\end{figure}

The reconstruction network correctly predicts the true microscopic distribution for all pulling velocities considered, with a mean log-likelihood of -1.23 on both the training set and the test set. The temporal evolution of the mean displacement is shown in Fig. \ref{fig:case1 IB mu sigma}, where an excellent agreement with the analytical evolution is observed. Similarly, the variance of the particle's displacement is approximately constant and independent of the pulling protocol, as expected from the analytical result (see figure in Appendix \ref{appx: case 1 std}). A single internal variable (as assumed in the chosen IB-VONNs architecture) is thus sufficient to capture the microscopic distribution. This learned variable $\alpha$ shall be a bijection of the particle's mean displacement as previously argued. To verify this claim, we depict in Fig. \ref{fig: case1 internal vs mean} the learned internal variable versus the true mean displacement of the training data. As may be observed, the data beautifully collapses onto a single monotonic function for all pulling velocities, hence confirming the capability of the proposed method in identifying a good internal variable that can accurately capture the relevant features of the microscopic distribution. The internal variables can then be used to sample microstates compatible with the macroscopic state of the system, providing some intuition to the meaning of the learned internal variable.

\begin{figure}[ht]
     \centering
     \begin{subfigure}{0.47\textwidth}
         \centering
         \includegraphics[width=\textwidth]{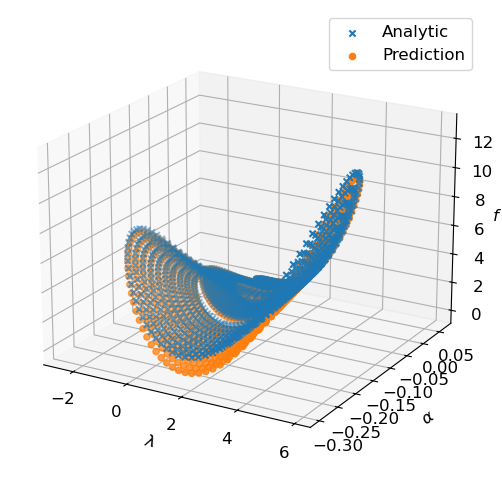}
         \caption{}
         \label{fig: case1_fe}
     \end{subfigure}
     \begin{subfigure}{0.47\textwidth}
         \centering
         \includegraphics[width=\textwidth]{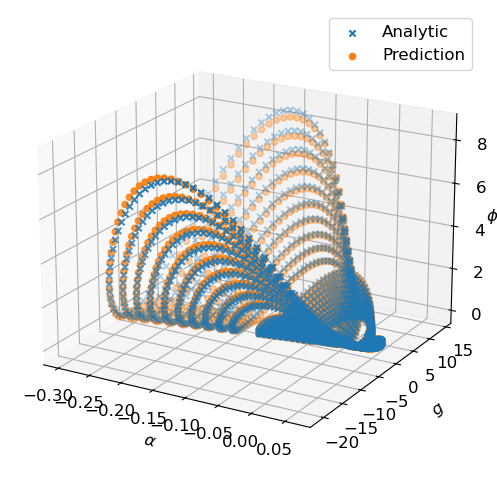}
         \caption{}
         \label{fig:case1_phi}
     \end{subfigure}
\caption{Comparison between the analytical values and the VONNs predictions of (a) the free energy density $f(\lambda,\alpha)$; (b) the dual dissipation potential density $\phi(\alpha,g)$ on the test set.}
\label{fig: case1_fe_psi}
\end{figure}

\begin{figure}[H]
     \centering
     \begin{subfigure}{0.40\textwidth}
         \centering
         \includegraphics[width=\textwidth]{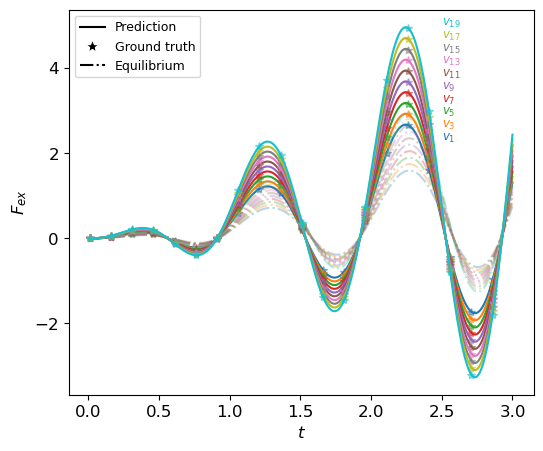}
         \caption{}
         \label{fig: case1_Fex}
     \end{subfigure}
     \begin{subfigure}{0.56\textwidth}
         \centering
         \includegraphics[width=\textwidth]{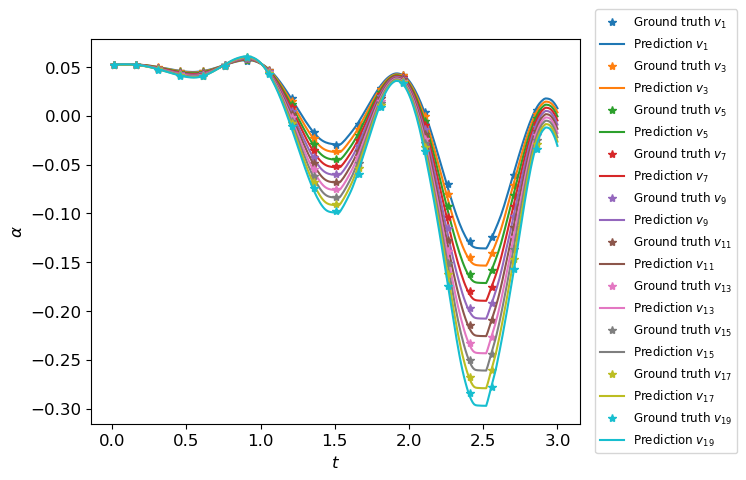}
         \caption{}
         \label{fig:case1_alpha}
     \end{subfigure}
\caption{Comparison between the test data and the VONNs predictions of (a) the external force $F_{ex}(t)$, and (b) the internal variable $\alpha(t)$ for all pulling $v_i$ in the test set. The equilibrium values of the external force are also plotted for all pulling velocities in order to quantify how far the system is driven away from equilibrium. The stars are the ground truth computed from Langevin simulations, and the solid lines are VONNs predictions.}
\label{fig: case1_Fex_alpha}
\end{figure}

The predicted thermodynamic potentials and their comparison to the analytical solution are depicted in Fig. \ref{fig: case1_fe_psi}. We quantify these predictions by means of the relative $L^2$ error, defined for a general vectorial function $\mathbf{A}(\mathbf{z})\in \mathbb{R}^n$ and its reference value $\mathbf{A}^{\text{ref}}(\mathbf{z})$ as 
\begin{equation} \label{eq: relative L2 error}
    \frac{\sqrt{\sum_{i=1}^n \int \left| A_i^{\text{ref}}(\mathbf{z}) - A_i(\mathbf{z}) \right|^2 \, d\mathbf{z} }}{\sqrt{\sum_{i=1}^n \int \left| A_i^{\text{ref}}(\mathbf{z}) \right|^2 d\mathbf{z}}} \times 100\%.
\end{equation}

The relative $L^2$ errors for $f$ and $\phi$ on the training set are $8.60\%$ and $4.99\%$, respectively. On the test set these are $8.60\%$ and $4.94\%$, respectively. As may be observed, there are noticeable yet reasonable deviations in both potentials, due to the fact that only the derivatives of such potentials are present in the loss function. With the learned potentials, one can then predict the evolution of the external force $F_{ex}(t)$ and the internal variable $\alpha(t)$. We do so for the pulling velocities used in the test set, and compare such evolutions to the ground truth in Fig.~\ref{fig: case1_Fex_alpha}. A remarkable accuracy is observed, despite small but noticeable errors in the thermodynamic potentials. The relative $L^2$ errors for $F_{ex}(t)$ and $\alpha(t)$ are $0.156\%$ and $2.69\%$, respectively, on the training set, and $0.171\%$ and $2.69\%$, respectively, on the test set. We note that the pulling protocol significantly pushes the system out of equilibrium. Indeed, the analytical predictions of the external force at equilibrium, shown in Fig.~\ref{fig: case1_Fex_alpha}(a), appreciably differ from the nonequilibrium values.

\begin{figure}[H]
    \centering
    \begin{subfigure}{0.47\textwidth}
         \centering
         \includegraphics[width=\textwidth]{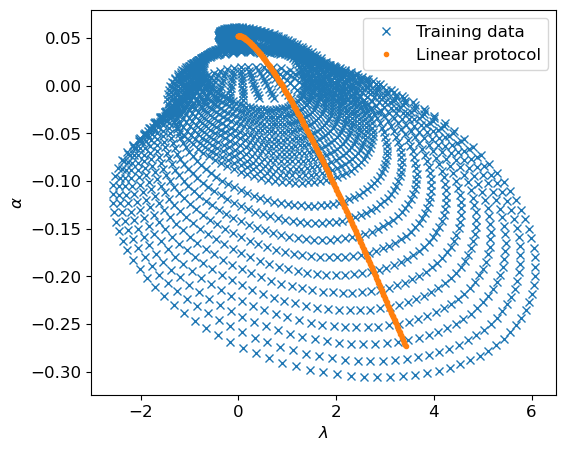}
         \caption{}
         \label{fig: case1 phase space}
    \end{subfigure}
    \begin{subfigure}{0.45\textwidth}
         \centering
         \includegraphics[width=\textwidth]{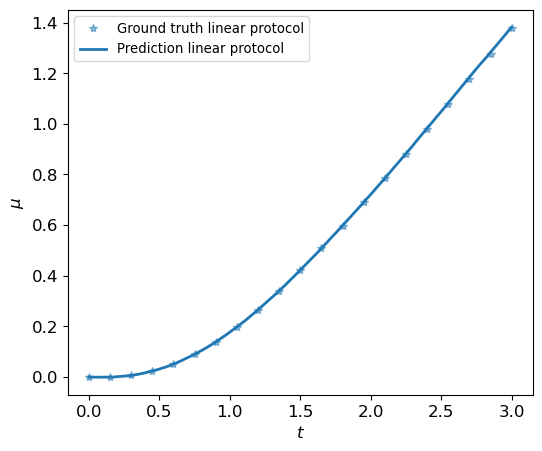}
         \caption{}
         \label{fig: case1 IB mu}
    \end{subfigure}
\caption{(a) Data distribution in the space of state variables $\alpha,\lambda$ for the training set as well as for the data generated by the smoothed linear pulling protocol. (b) Prediction of the mean displacement versus the ground truth (directly computed from the Langevin simulations) for the smoothed linear pulling protocol. }
\end{figure}

\begin{figure}[H]
     \centering
     \begin{subfigure}{0.45\textwidth}
         \centering
         \includegraphics[width=\textwidth]{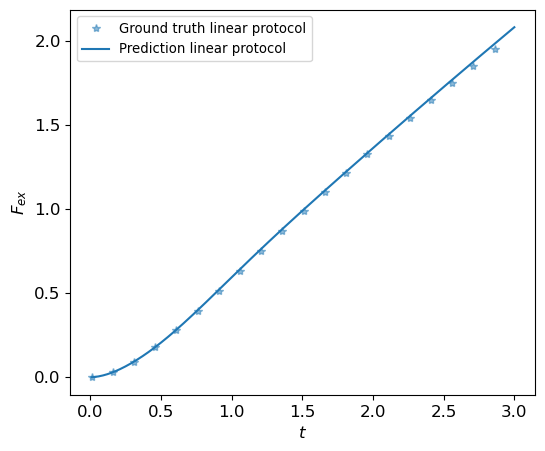}
         \caption{}
         \label{fig: case1_Fex_lin}
     \end{subfigure}
     \begin{subfigure}{0.47\textwidth}
         \centering
         \includegraphics[width=\textwidth]{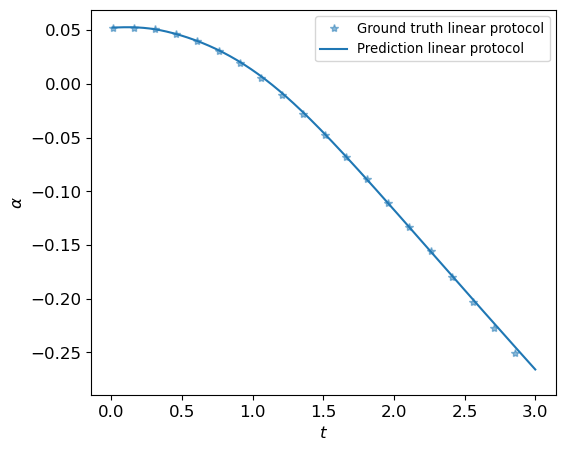}
         \caption{}
         \label{fig:case1_alpha_lin}
     \end{subfigure}
\caption{Comparison between the data generated by a smoothed linear pulling protocol $\lambda(t)=1.2t(1 - e^t)$ and the VONNs predictions of the dynamics of (a) the external force $F_{ex}(t)$; (b) the internal variable $\alpha(t)$. The stars are the ground truth computed from Langevin simulations, and the solid lines are VONNs predictions.}
\label{fig: case1_Fex_alpha_lin}
\end{figure}

To further demonstrate the capability of the model, we also test the IB-VONNs famework on data generated by a smoothed linear pulling protocol $\lambda(t)=1.2t(1 - e^t)$. This has a different functional form from the protocols used to generate the training set, though the mean particle position still lies within the range of the training set (Fig. \ref{fig: case1 phase space}). As shown in Fig.s~\ref{fig: case1 IB mu} and \ref{fig: case1_Fex_alpha_lin}, both networks work very well for the new pulling protocol, with a relative $L^2$ errors of $1.20\%$ and $1.77\%$ for $F_{ex}(t)$ and $\alpha(t)$, respectively, thus giving empirical evidence of the generalizabilty of the proposed strategy. 

\section{Example 2: a mass-spring chain with double-well potential} \label{sec: case2}
\subsection{Model setup and simulation details} \label{sec: example2_setup}
We consider a more complicated model in this section, namely, a one-dimensional mass-spring chain where masses interact through a double-well potential. This model has been widely used to study phase transitions \cite{rose2004scaffolds,torres2019combined} and hysteresis phenomena \cite{fraternali2011multiscale,puglisi2005thermodynamics}, and, despite its apparent simplicity, it is of marked theoretical difficulty. Indeed, a statistical mechanics and thermodynamics foundation for this system has been elusive till recently \cite{leadbetter2023statistical,leadbetter2024statistical}, where only the low temperature regime has been successfully coarse-grained for arbitrary pulling protocols. In this regime the probability distribution is approximately a multivariate Gaussian, except near transition points. Here, we consider instead a higher temperature regime, where the distribution is markedly multi-peaked at all times. 

\begin{figure}[ht]
    \centering
    \includegraphics[width=0.4\textwidth]{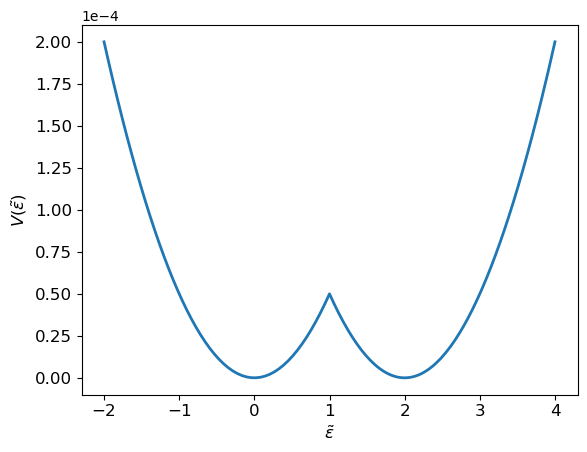}
    \caption{The interparticle potential in Example 2.}
    \label{fig: case2 potential}
\end{figure}

In the following, we consider a mass spring chain with $n = 10$ particles, where the leftmost particle is attached to a fixed wall via a spring, and the rightmost particle is pulled according to a prescribed displacement protocol $x_{10}=\lambda(t)$. The system is initialized according to the equilibrium distribution for the given initial streth $\lambda(0)$, and then driven out of equilibrium through the external protocol. 
The displacements and velocities of the particles are denoted by $x_i(t)$ and $\dot{x}_i(t)$, respectively, with $i=1,\cdots,10$, and the strain of each spring is defined as $\tilde{\varepsilon}_i(t) = (x_i(t) - x_{i-1}(t)) / L_0$, where $L_0$ is the initial length of the springs and $x_0=0$ (the symbol $\varepsilon_i$ will be reserved for the local macroscopic strain value). The system obeys overdamped Langevin dynamics with nearest-neighbor interaction
\begin{equation} \label{Eq:Langevin_case2}
    \eta \dot{x}_i = -\frac{\partial V(\tilde{\varepsilon}_i)}{\partial x_i} - \frac{\partial V(\tilde{\varepsilon}_{i+1})}{\partial x_i} + \sqrt{2\eta k_B T} \dot{\xi}_i,
\end{equation}
where $\eta$ is the viscosity, $k_B$ the Boltzmann constant, $T$ the temperature and $\dot{\xi}_i$ is a white noise. The interparticle potential $V$ is taken to be of the form
\begin{equation}
    V(\tilde{\varepsilon}_i) = \begin{cases}
  \frac{1}{2} E_1 L_0 \tilde{\varepsilon}_i^2, & \tilde{\varepsilon}_i \leq {\varepsilon}_c, \\
  \frac{1}{2} E_2 L_0 (\tilde{\varepsilon}_i - {\varepsilon}_h)^2, & \tilde{\varepsilon}_i > {\varepsilon}_c,
\end{cases}
\end{equation}
see Fig.~\ref{fig: case2 potential}, where $E_1, E_2, \varepsilon_c, \varepsilon_h$ are constants satisfying $V(\varepsilon_c -) = V(\varepsilon_c +)$, or equivalently, $E_1 \varepsilon_c^2 = E_2 (\varepsilon_c - \varepsilon_h)^2$. 

Similarly to the previous example, $x_i(t)$ and $\dot{x}_i(t)$ are realizations of the stochastic dynamics, and hence microscopic in nature. In contrast, the (macroscopic) state variables are defined via ensemble averages. These consist of the mean strain for each spring $\varepsilon_i(t)=\left< \tilde{\varepsilon}_i(t) \right>$,  and the internal variables $\alpha_i$ we are aiming to discover, i.e., $\boldsymbol{z}_i = (\varepsilon_i, \alpha_i)^T$. We assume that there is only one internal variable per spring. The natural (macroscopic) process variables for this system are thus the mean velocity $v_i(t)=\left< \dot{x}_i(t) \right>$, and the rate of change of the internal variable $\dot{\alpha}_i$, i.e., $\boldsymbol{w}_i = (v_i, \dot{\alpha}_i)^T$. These will be of relevance for deriving the functional form of the dynamics through Onsager's variational principle, which we will do in the next section.

The Langevin dynamic equations, Eq.~\eqref{Eq:Langevin_case2}, are simulated with parameters
\begin{equation}
    \eta = 0.1, \quad E_1 = E_2 = 10^{-2}, \quad u_c = 1, \quad u_h = 2, \quad L_0 = 0.01, \quad k_B T = 5\times 10^{-6}
\end{equation}
using the Euler-Maruyama scheme for $t\in[0,3]$ using an equally-spaced time grid with a time step $\Delta t=0.01$. The system is first held fixed at a total length $L_{tot}$ and equilibrated for the prescribed temperature. Then, the last particle is pulled by a smoothed linear protocol $\lambda(t) = vt(1 - e^{-t})$, where $v$ is set to be $0.05$. To explore the phase space of the state variables, instead of considering different pulling protocols, we consider different initial conditions, which turns out to work better in this case. 17 initial conditions are used with $L_{tot}$ ranging from $0.12$ to $0.28$ with equal spacing, i.e, $L_{tot,i}=0.11+0.01i, i=0,1,\cdots,16$, and $n_r = 2,000$ trajectories are sampled under each initial condition. The data generated with initial total lengths $L_{tot,0}, L_{tot,2}, \cdots, L_{tot,16}$ is used as the training set, and the rest of the data is used as the test set.

A representative marginal probability distribution of the various spring strains is represented in Fig. \ref{fig: IB_GMM_pdf} in Appendix \ref{Appx: case 2 IB_GMM} for a given initial length and intermediate time. For the chosen temperature and pulling protocol, some of these distributions are clearly distinct from the equilibrium distribution (also shown in the figure), and hence cannot be described by the local strain exclusively. We here use one internal variable in addition to the mean strain for each spring to describe the state of the system. Although more internal variables could of course be used, one per spring will be sufficient to obtain a first order approximation of the macroscopic evolution of the system (see results in Section \ref{Sec:case2_IB_VONNs} and \ref{sec: case 2 IB_CNF}).

\subsection{Evolution equations of the state variables} \label{sec: case2 evl eqn}
We here aim at recovering (spatially-discretized) continuum evolution equations, for which it is typically assumed  the existence of free energy and dissipation potential densities. Similarly, we here consider that $\mathcal{F}$ and $D$ take the form
\begin{equation}
    \mathcal{F} = \sum_{i=1}^{n} f_i, \quad D = \sum_{i=1}^{n} \psi_i,
\end{equation}
where, in this case, we consider
\begin{equation}
    f_i = f(\varepsilon_i, \alpha_i), \quad \psi_i = \psi(\varepsilon_i, \alpha_i, v_i,\dot{\alpha}_i).
\end{equation}
We remark that, inspired by \cite{leadbetter2023statistical}, we have not included in $f$ the standard gradient term used in phase field models \cite{chen2002phase} (we note that $\alpha$ does not need to take the meaning of the fraction in a given phase). This spatially local assumption will result in a specific structure of the evolution equations, which, in turn, will guide the architecture of the VONNs method. 
In particular, by Onsager's variational principle, the Rayleighian is defined as $\mathcal{R} = \dot{\mathcal{F}} + \mathcal{D} + \mathcal{P}$, where the external power is $\mathcal{P} = -F_{ex} v_{10}$, with $F_{ex}$ being the ensemble average of the external force exerted on the last particle. The evolution equations read
\begin{align} 
    \label{eq: case2 eps} \frac{1}{L_0} \left(\frac{\partial f_{i+1}}{\partial \varepsilon_{i+1}} - \frac{\partial f_i}{\partial \varepsilon_{i}} \right) &= \frac{\partial \psi_i}{\partial v_i}, \quad i=1,2,\cdots,9,  \\
    \label{eq: case2 f bc} F_{ex} - \frac{1}{L_0} \frac{\partial f_{10}}{\partial \varepsilon_{10}} &= \frac{\partial \psi_{10}}{\partial v_{10}}, \\
    \label{eq: case2 alpha} - \frac{\partial f_i}{\partial \alpha_{i}} &= \frac{\partial \psi_i}{\partial \dot{\alpha_i}}, \quad i=1,\cdots, 10.
\end{align}



\begin{figure}[t]
    \centering
    \includegraphics[width=0.9\textwidth]{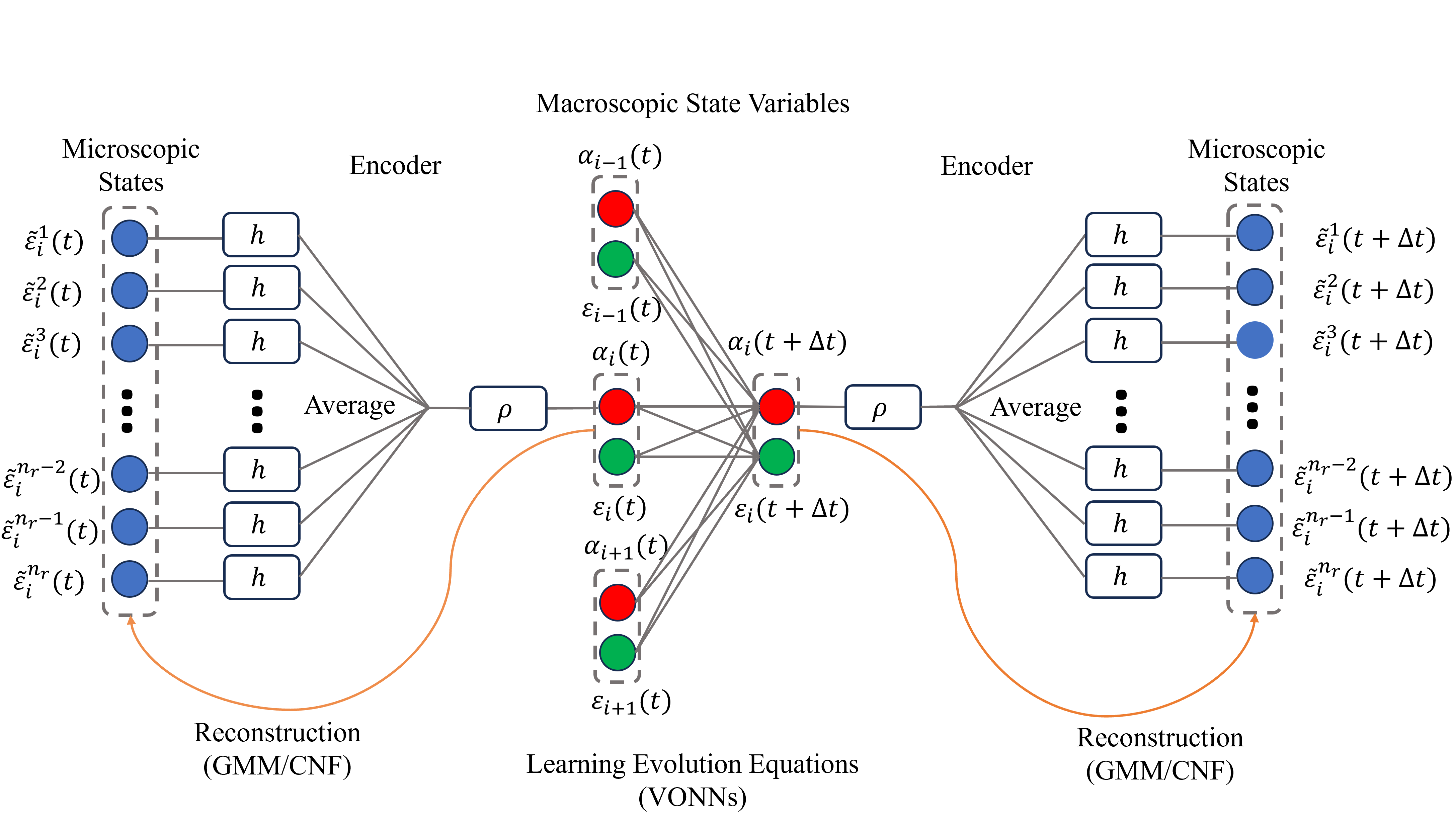}
    \caption{A schematic plot of the IB-VONNs architecture. As shown in the middle part of the figure, the evolution of the state variables depends on that of the same location as well as that of the nearest neighbors.}
    \label{fig: case 2 IB-CNF}
\end{figure}

In practice, we use the formulation of VONNs written in terms of the free energy and dual dissipation potential (see Section \ref{Sec:VONNs}) to learn the evolution equations of the state variables. We model the free energy density $f$ by an INN and the dual dissipation potential density $\phi$ by a PICINN. The derivatives of the dissipation potential with respect to the process variables are not directly measurable from data, and are hence calculated from the free energy densities from Eqs.~\eqref{eq: case2 eps} - \eqref{eq: case2 alpha}, i.e.,
\begin{align} 
     g_{v_i} &\coloneqq \frac{\partial \psi_i}{\partial v_i} = \frac{1}{L_0} \left(\frac{\partial f_{i+1}}{\partial \varepsilon_{i+1}} - \frac{\partial f_i}{\partial \varepsilon_{i}} \right), \quad i=1,2,\cdots,9,  \\ \label{Eq:g_10_Fex}
     g_{v_{10}} &\coloneqq \frac{\partial \psi_{10}}{\partial v_{10}} = F_{ex} - \frac{1}{L_0} \frac{\partial f_{10}}{\partial \varepsilon_{10}}  \\
     g_{\dot{\alpha}_i} &\coloneqq \frac{\partial \psi_i}{\partial \dot{\alpha}_i} = - \frac{\partial f_i}{\partial \alpha_{i}}, \quad i=1,\cdots, 10.
\end{align}
These will be inputs to the dual dissipation potential density, naturally inducing a spatially non-local dependence in the evolution equations. That is, the future state $\varepsilon_i(t+\Delta t), \alpha_i(t+\Delta t)$ will not only depend on $\varepsilon_i(t), \alpha_i(t)$, but also on $\varepsilon_{i+1}(t), \alpha_{i+1}(t)$ and $\varepsilon_{i-1}(t), \alpha_{i-1}(t)$, see Fig.~\ref{fig: case 2 IB-CNF}. The evolution equations now read
\begin{equation} \label{eq: legendre vonns v}
     v_i =\frac{\partial \phi_i}{\partial g_{v_i}}\left(\varepsilon_i, \alpha_i, g_{v_i}, g_{\dot{\alpha}_i} \right),
\end{equation}
\begin{equation} \label{eq: legendre vonns alpha dot}
     \dot{\alpha}_i = \frac{\partial \phi_i}{\partial g_{\dot{\alpha}_i}}\left(\varepsilon_i, \alpha_i, g_{v_i}, g_{\dot{\alpha}_i} \right).
\end{equation}
Note that, by definition, we have the identity $\dot{\varepsilon}_i=\frac{v_i-v_{i-1}}{L_0}$ (see Section \ref{sec: example2_setup}), which can be used to obtain the evolution equation for $\varepsilon_i$. The values of $\varepsilon^{NN}_i(s)$ and $\alpha^{NN}_i(s)$ can then be obtained via numerical integration, i.e.,  
\begin{equation}
    \varepsilon_i^{NN}(s) = \int_{t_0}^{s} \dot{\varepsilon}^{NN}_i(s') ds',
\end{equation}
\begin{equation}
    \alpha_i^{NN}(s) = \int_{t_0}^{s} \dot{\alpha}_i(s')^{NN} ds',
\end{equation}
where we use a forward Euler scheme. 

The neural networks associated to the encoder-decoder structure and VONNs will be jointly trained using the loss function
\begin{align}
    \mathcal{L} =& -\frac{\lambda_{pdf}}{N_{pdf}} \sum_{i=1}^n \sum_{j=0}^{n_t} \sum_{k=1}^{n_r} \log q(\tilde{\varepsilon}_i^k(t_0 + j\cdot \Delta t)|\boldsymbol{z}_i(t_0 + j\cdot \Delta t)) \\
    &+ \frac{\lambda_{\varepsilon}}{N} \sum_{i=1}^n \sum_{j=1}^{n_t} \Big| \varepsilon^{NN}_i(t_0 + j\cdot \Delta t) - \varepsilon_i(t_0 + j\cdot \Delta t) \Big|^2\\
    &+ \frac{\lambda_{\alpha}}{N} \sum_{i=1}^n \sum_{j=1}^{n_t} \Big| \alpha^{NN}_i(t_0 + j\cdot \Delta t) - \alpha_i(t_0 + j\cdot \Delta t) \Big|^2.
\end{align}
Here, $N_{pdf}=n(n_t+1)n_r$ and $N=nn_t$ are the numbers of training data points for the various loss terms.

In the training process, the measurement of the external force $F_{ex}$ is used as an input (note that it appears in Eq.~\eqref{Eq:g_10_Fex}). For predictions though, one may either consider the system with either traction boundary conditions ($F_{ex}$ known, $v_{10}$ unknown) or displacement boundary conditions ($v_{10}$ known, $F_{ex}$ unknown). In the first scenario, predicting the evolution of the state variables is straightforward, while in the second scenario, an additional step to predict $F_{ex}$ is needed. In that case, $F_{ex}$ is modeled by a neural network with $\varepsilon_{10}, {\alpha_{10}}$ and $v_{10}$ as input and it is updated iteratively at each time step. Note that here $\dot{\alpha}_{10}$ is not necessary to predict $F_{ex}$, since we know from Eqs.~\eqref{eq: case2 alpha} that $\dot{\alpha}_{10}$ can be predicted from $\varepsilon_{10}, {\alpha_{10}}$ and $v_{10}$.

\subsection{Learning of the internal variables and evolution equations through the IB-VONNs method with GMM} \label{Sec:case2_IB_VONNs}
In this section, as a first step, we model the marginals of the microscopic distribution with a Gaussian Mixture Model (GMM). This assumption will be relaxed in the following section, where the more general IB-VONNs method with CNFs will be used to jointly learn the internal variables and their dynamics, with no a priori assumption on the microscopic distribution. 

The microscopic marginal distribution of the strain of each spring is modeled with a GMM composed of two Gaussians as (the subscripts "i" to index the springs are omitted for simplicity)
\begin{equation}
     q(\tilde{\varepsilon}(t)|\boldsymbol{z}(t)) = \pi(\alpha(t)) \mathcal{N}\left(\varepsilon (t) + (\pi(\alpha(t))- 1) \varepsilon_h, \sigma_l^2(\boldsymbol{z}(t))\right) + (1 - \pi(\alpha(t))) \mathcal{N}\left(\varepsilon(t) + \pi(\alpha(t)) \varepsilon_h, \sigma_r^2(\boldsymbol{z}(t))\right),
\end{equation}
where $\pi(\alpha(t))$ is the mixing coefficient, $\varepsilon$ corresponds to the mean strain, and $\sigma_l^2(\boldsymbol{z}(t))$ and $\sigma_r^2(\boldsymbol{z}(t))$ are the variances of the two Gaussians. The "distance" between the peaks of the two Gaussians is assumed to be given by $\varepsilon_h$, which corresponds to the strain between the two wells of the interparticle potential. This assumption will turn out to be quite accurate in the examples here considered. Here, $\pi$ is modeled by a neural network with internal variable $\alpha(t)$ as input, and $\sigma_l^2$ and $\sigma_r^2$ are modeled by the two neural networks with the state variables $\boldsymbol{z}(t)$ as input (see Fig. \ref{fig: GMM}). These neural networks are composed of 2 hidden layers and 10 neurons per layer.  

The architecture of the encoder, used to calculate the internal variables (see Section \ref{sec: vib}), is as follows. We use a neural network with 2 hidden layers with 10 neurons to model the function $h$, and a neural network with 1 hidden layer with 10 neurons to model the nonlinear transformation $\rho$. As for the VONNs part, an INN with 2 layers and 20 neurons per layer is used to approximate the free energy density, and a PICINN with 2 layers and 20 neurons per layer for each, the convex and non-convex branch, is used to approximate the dual dissipation potential. The characteristic scales are estimated based on Eq.~\eqref{eq: case2 eps} and \eqref{eq: case2 alpha} as
\begin{equation}
    f^{*} = \sigma_{F_{ex}} \sigma_{\varepsilon} L_0, \quad \phi^{*} = f^{*},
\end{equation}
where $\sigma_{F_{ex}}$ and $\sigma_{\varepsilon}$ stand for the standard deviation of the external force and mean strain, respectively, and $L_0$ is the initial length of the springs. The loss weights are chosen empirically and found to be crucial for the success of the training. $\lambda_{\varepsilon}$ and $\lambda_{\alpha}$ are set to be 1, and $\lambda_{pdf}$ is set to 0.002. Mini-batch training is adopted to reduce memory consumption and to avoid local minima. A batch consists of data from one of the initial conditions (randomly chosen) within the time range of $[t_0, t_f]$, where the starting time $t_0$ is chosen randomly from $[0,3-(t_f-t_0)]$. $t_f-t_0$ is set to $0.3$ initially, and increased to $3$ during the training process. The learning rate is set to $5 \times 10^{-5}$.

As shown in Figure \ref{fig: IB_GMM_pdf} in Appendix \ref{Appx: case 2 IB_GMM}, the learned internal variables can predict the microscopic distributions reasonably well (note that such distributions are distinct from the equilibrium one, also represented for comparison in the same figure).  The mean log-likelihood is 0.517 on the training set and 0.589 on the test set. This ensures that the learned internal variables together with the mean local strains are capable of capturing the important features of the microscopic distribution. The data distribution in the phase space ($\alpha-\varepsilon$ space) for all initial conditions considered is shown in Fig. \ref{fig: IB_GMM_phase_space}. For a mean strain $\varepsilon$ below 3, the distribution is multimodal, and the internal variable characterizes that feature. For $\varepsilon>3$ the distribution is single peaked and close to a Gaussian distribution, and can hence be described by the mean strain exclusively (as long as the variance remains relatively constant). In this case, the data points collapse into a single curve, and the internal variable effectively becomes a function of the mean strain. 



Since we do not have analytic solutions for the free energy and dissipation potential in this case, we quantify the accuracy of the results by the relative $L^2$ error, as defined in Eq.~\eqref{eq: relative L2 error}, of the mean strain and the internal variable. In the first scenario, we investigate the performance of our method with known external force $F_{ex}$. As shown in Fig.~\ref{fig: IB_GMM_eps_Fex_input} and \ref{fig: IB_GMM_alpha_Fex_input}, the predictions by VONNs are in good agreement with the ground truth, with mean relative $L^2$ errors on the training set (over all positions and initial conditions) of $1.22\%$ and $2.90\%$ for the mean strain and internal variable, respectively. Similarly, the mean relative $L^2$ errors on the test set of the mean strain and internal variable are $1.37\%$ and $2.48\%$, respectively. This verifies that the learned internal variables together with the local strain values can predict the dynamics without using the history as input. In the second scenario, we investigate the performance of our method with known velocity of the last particle $v_{10}$, as shown in Fig.~\ref{fig: IB_GMM_Fex} -~\ref{fig: IB_GMM_alpha_v_input}. The mean relative $L^2$ errors on the training set are $0.960\%$, $2.74\%$ and $2.24\%$ for the mean strain, internal variable and external force, respectively. Similarly, the mean relative $L^2$ errors on the test set for the mean strain, internal variable and external force are $1.41\%$, $2.43\%$ and $2.18\%$. We also investigate the generalizability of the method by testing it on data generated by the unseen sinusoidal pulling protocol $\lambda(t) = vt(1 - e^t)(1+0.2\sin(\pi t))$. Even though the model is trained with only one protocol, it still performs reasonably well on this new protocol, see Fig. \ref{fig: IB_GMM_eps_alpha_v_input_sin}. With known external force as input, the mean relative $L^2$ errors of the mean strain and internal variable are $4.42\%$ and $6.05\%$, respectively. With known velocity of the last particle as input, the mean relative $L^2$ errors of the mean strain, internal variable and external force are $6.66\%$, $6.40\%$ and $16.0\%$, respectively. 

\subsection{Learning of the internal variables and evolution equations through the IB-VONNs method with CNFs} \label{sec: case 2 IB_CNF}

The GMM used in the previous section was motivated by physical intuition. However, such physical insight may not be always available under far-from-equilibrium conditions, especially for high-dimensional problems. We here test the IB-VONNs framework with CNFs proposed in Section \ref{sec: CNF}, which makes no a priori assumptions on the form of the microscopic distribution. To enable an evaluation of its performance, we consider the same system and data set as that of the previous section. 

We use the same architecture as in the previous section for the encoder (i.e., function $h$ and $\rho$) to learn the internal variables as well as for VONNs to learn the evolution equations. The parameters $\boldsymbol{a}, \boldsymbol{b}$ and $\boldsymbol{c}$ in the CNFs are modeled by the three outputs of a shared neural network with 2 hidden layers and 10 neurons per layer, and $d$ is set to be 2; see Fig.~\ref{fig: CNF}. We add an exponential transform to guarantee the positivity of $\boldsymbol{a}$, and use a softmax function to enforce the constraints on $\boldsymbol{c}$. The loss weights are chosen empirically. $\lambda_{\varepsilon}$ and $\lambda_{\alpha}$ are set to be 1, and $\lambda_{pdf}$ is set to be 0.001. Same mini-batch training strategy is adopted as in the previous section. 

\begin{figure}[H]
     \centering
     \begin{subfigure}{0.40\textwidth}
         \centering
         \includegraphics[width=\textwidth]{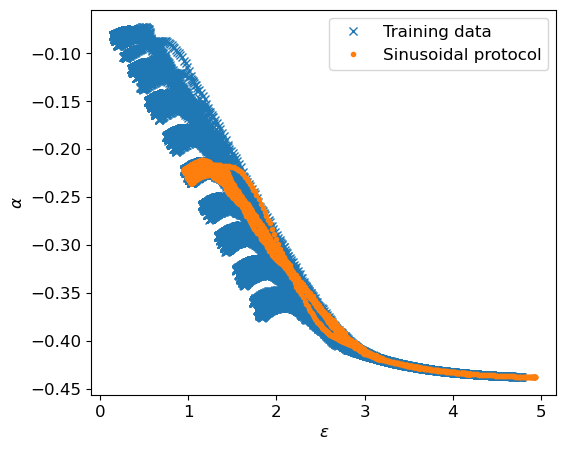}
         \caption{}
         \label{fig: IB_CNF_phase_space}
     \end{subfigure}
     \begin{subfigure}{0.54\textwidth}
         \centering
         \includegraphics[width=\textwidth]{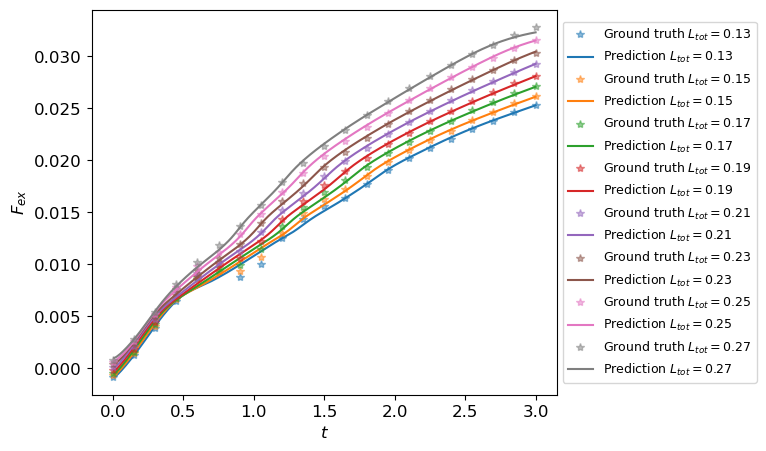}
         \caption{}
         \label{fig: IB_CNF_Fex}
     \end{subfigure}
     
     \centering
     \begin{subfigure}{0.43\textwidth}
         \centering
         \includegraphics[width=\textwidth]{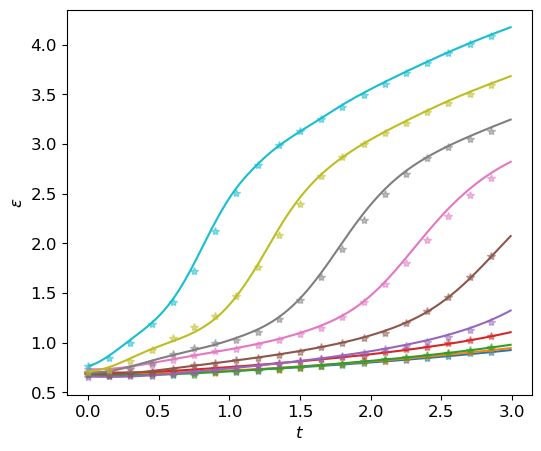}
         \caption{}
         \label{fig: IB_CNF_eps_v_input}
     \end{subfigure}
     \begin{subfigure}{0.53\textwidth}
         \centering
         \includegraphics[width=\textwidth]{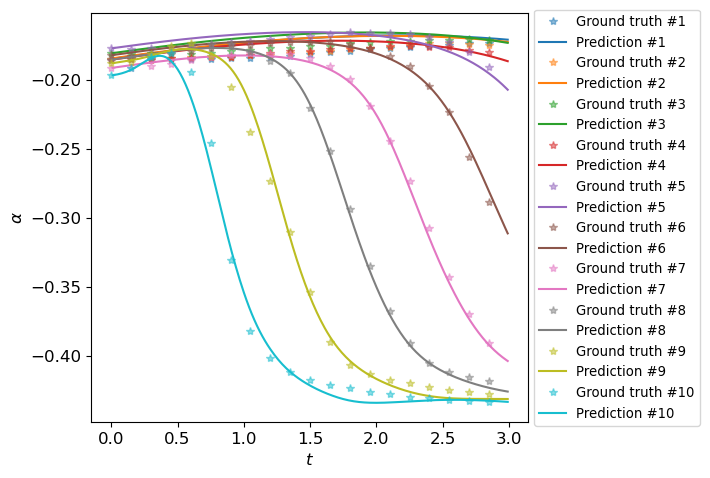}
         \caption{}
         \label{fig: IB_CNF_alpha_v_input}
     \end{subfigure}
\caption{Results from the IB-VONNs method with CNFs. (a) The distribution of training data and the data generated by sinusoidal pulling protocol in the phase space of the state variables; (b) Predictions of the external force on the test set; Predictions of (c) the mean stain and (d) internal variable on data with initial condition $L_{tot}=0.17$ (test set). In (b) - (d), velocity of the last particle is fed to the model as input. The stars in (b) - (d) are ground truth computed from Langevin simulations, and the solid lines are VONNs predictions. Numbers in the legends are the indices of the springs.}
\label{fig: IB_CNF_phase_Fex_eps_alpha_v_input}
\end{figure}

The marginals of the learned miscroscopic distribution $q(\tilde{\varepsilon}_i(t)|\boldsymbol{z}_i(t))$ are shown in Fig.~\ref{fig: IB_CNF_pdf} of Appendix \ref{Appx: case 2 IB_CNF} for an intermediate time together with the true distribution. As may be there observed, the flexibility from the CNFs framework enables a more accurate representation of the probability distribution as compared to the GMM. The mean log-likelihood is -0.269 on the training set and -0.337 on the test set, which are indeed significantly lower. We remark that the added flexibility of the CNFs is not done at the expense of an increased complexity of the macroscopic model. Here, a single internal variable per spring is also used. The data distribution in the phase space ($\alpha-\varepsilon$ space) for the training set is shown in Fig.~\ref{fig: IB_CNF_phase_space}. Although the range of the learned internal variable differs from the results of Section \ref{Sec:case2_IB_VONNs}, the distribution is qualitatively very similar to that of Fig.~\ref{fig: IB_GMM_phase_space}: below $\varepsilon<3$, when the distribution is multimodal, the internal variable provides additional information to the mean strain, while for $\varepsilon>3$, the internal variable effectively becomes a function of $\varepsilon$. 

\begin{figure}[H]
     \centering
     \begin{subfigure}{0.42\textwidth}
         \centering
         \includegraphics[width=\textwidth]{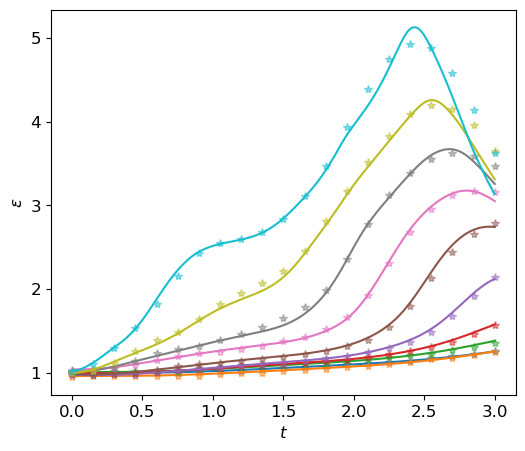}
         \caption{}
         \label{fig: IB_CNF_eps_v_input_sin}
     \end{subfigure}
     \begin{subfigure}{0.54\textwidth}
         \centering
         \includegraphics[width=\textwidth]{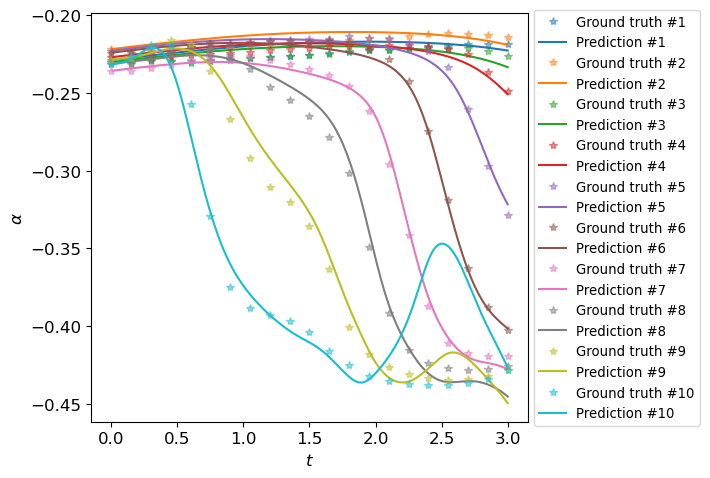}
         \caption{}
         \label{fig: IB_CNF_alpha_v_input_sin}
     \end{subfigure}
    \begin{subfigure}{0.54\textwidth}
         \centering
         \includegraphics[width=\textwidth]{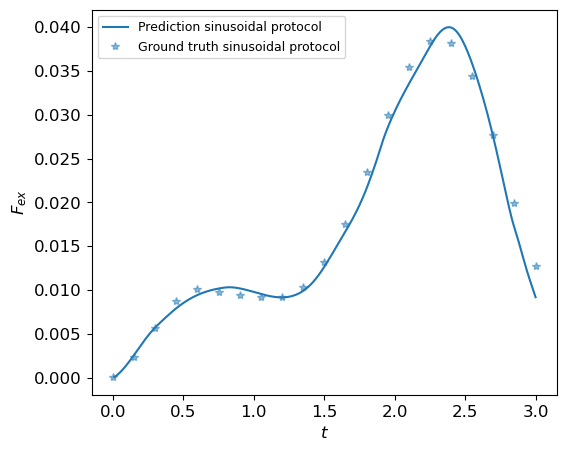}
         \caption{}
         \label{fig: IB_CNF_Fex_sin}
     \end{subfigure}
\caption{IB-VONNs with CNFs model tested on data with initial condition $L_{tot}=0.20$ and sinusoidal pulling protocol $\lambda(t)=vt(1 - e^t)(1+0.2\sin(\pi t))$. Predictions of (a) the mean strain, (b) the internal variable, and (c) external force with the velocity of the last particle as input. The stars are ground truth computed from Langevin simulations, and the solidlines are VONNs predictions. Numbers in the legends are the indices of the springs.}
\label{fig: IB_CNF_eps_alpha_v_input_sin}
\end{figure}

IB-VONNs with CNFs also performs reasonably well in predicting the dynamics of the state variables and the external force. In the scenario of known external force $F_{ex}$, as shown in Fig.~\ref{fig: IB_CNF_eps_Fex_input} and \ref{fig: IB_CNF_alpha_Fex_input}, the predictions by VONNs are in good agreement with the ground truth, with mean relative $L^2$ errors on the training set (over all positions and initial conditions) of $1.06\%$ and $2.16\%$ for the mean strain and internal variable, respectively. Similarly, the mean relative $L^2$ errors on the test set of the mean strain and internal variable are $0.994\%$ and $2.12\%$, respectively. In the second scenario of known velocity of the last particle $v_{10}$, as shown in Fig.~\ref{fig: IB_CNF_Fex} -~\ref{fig: IB_CNF_alpha_v_input}, the mean relative $L^2$ errors on the training set are $0.720\%$, $1.97\%$ and $1.85\%$ for the mean strain, internal variable and external force, respectively. Similarly, the mean relative $L^2$ errors on the test set for the mean strain, internal variable and external force are $0.848\%$, $1.99\%$ and $1.66\%$. We also investigate the generalizability of the method by testing on data generated by unseen sinusoidal pulling protocol $\lambda(t) = vt(1 - e^t)(1+0.2\sin(\pi t))$. Even though the model is trained with only one protocol, it still performs reasonably well on this new protocol, see Fig. \ref{fig: IB_CNF_eps_alpha_v_input_sin}. With known external force as input, the mean relative $L^2$ errors of the mean strain and internal variable are $2.67\%$ and $4.43\%$, respectively. With known velocity of the last particle as input, the mean relative $L^2$ errors of the mean strain, internal variable and external force are $3.35\%$, $4.31\%$ and $6.00\%$, respectively. We remark that IB-VONNs with CNFs outperforms IB-VONNs with GMM in accuracy and generalizability because of the capability of CNFs to learn arbitrary microscopic distributions.


\section{Conclusions} \label{sec: conclusions}
In this work we propose IB-VONNs, a new framework to learn internal variables and their evolution equations, consistently with the principles of non-equilibrium statistical mechanics and thermodynamics. The internal variables are learned through an encoder-decoder structure based on the IB principle that guarantees that the internal variables are macroscopic in nature, invariant with respect to the order of realizations, and can describe, jointly with the state variables used in equilibrium, the microscopic probability distribution at all times (such distribution will differ from the Boltzmann distribution away from equilibrium, and can be represented, in general, with conditional normalizing flows). This autoencoder structure is jointly trained with VONNs to ensure that the learned state variables lead to Markovian dynamics that are compliant with the laws of thermodynamics. In this work we model the free energy and dual dissipation potential (as opposed to the dissipation potential in the original formulation for VONNs), which turns out to be highly convenient for their integration with other tasks.


We investigate the performance of the proposed method on two problems governed by overdamped Langevin dynamics. The first one is a prototypical model for a particle in an optical trap, where the evolution of the true probability distribution (Gaussian at all times) and the associated free energy and dissipation potential may be computed analytically, thus serving as an ideal test bed for the proposed approach. The second example is a one-dimensional mass-spring chain at a relatively high temperature with a double-welled interaction potential. This problem, despite its apparent simplicity, is characterized by a multi-peaked probability distribution, and lacks, to this date, an ab initio derivation of its macroscopic description. In both cases, IB-VONNs was successful in discovering meaningful internal variables, and learn evolution equations that were predictive of macroscopic observables, such as the external driving force. 
 

These results obtained provide empirical evidence of the capability of the proposed approach in bridging statistical mechanics and thermodynamics away from equilibrium. Several interesting open questions remain to be addressed in the future though. One of these questions is how to quantify the uncertainty in the predictions and increase the robustness of the framework with very noisy data. We remark that the microscopic data in this work is intrinsically noisy since it is generated by stochastic dynamics (this would also be the case in Hamiltonian dynamics due to thermal fluctuations). In the current strategy, we have integrated the derivatives of the state variables in the expression of the loss function to make the model more robust, though this is done at the expense of an increased computational cost. Another technical problem is how to automatically determine the loss weights to increase the robustness of the learning strategy. 
Finally, the application to higher-dimensional systems is also of interest. These and other efforts will be the goal of future investigations. 

\appendix
\newpage
\section*{Acknowledgments}
The authors acknowledge support from NSF CMMI-2047506.

\section{Further details on example 1} \label{appx: case 1}
\subsection{Standard deviation of the particle displacement} \label{appx: case 1 std}
\begin{figure}[H]
  \centering
  \includegraphics[width=0.5\textwidth]{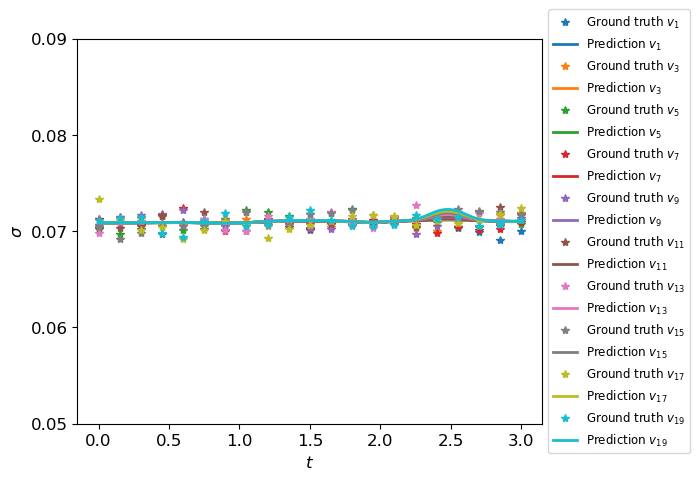}
  \caption{Predictions of the standard deviation of the displacement by the decoder versus the ground truth for the pulling velocities $v_i$ considered in the test set in example 1. The standard deviation remains approximately constant (with fluctuations) during the process, in agreement with the analytical predictions.}
  \label{fig: case 1 std}
\end{figure}

\subsection{Uniqueness of the free energy and dissipation potential}
We here prove, by contradiction, that the inverse problem of identifying the free energy and (dual) dissipation potential densities from the dynamics is unique for example 1. Let $f^{(i)}(\lambda,\alpha), \psi^{(i)}(\alpha,\dot\alpha) \, (i=1,2)$ be two solutions to Eqs.~\eqref{eq: case1 OVP fex}-\eqref{eq: case1 OVP iv}, and consider the difference $\Delta f = f^{(1)} - f^{(2)}, \, \Delta \psi = \psi^{(1)} - \psi^{(2)}$. Then, their governing equations read
\begin{align} 
&\frac{\partial \Delta f}{\partial \lambda} = 0, \label{eq: case1 OVP delta fex}\\ 
&\frac{\partial \Delta f}{\partial \alpha} + \frac{\partial \Delta \psi}{\partial \dot{\alpha}} = 0  \label{eq: case1 OVP delta iv} .
\end{align}
From Eq.~\eqref{eq: case1 OVP delta fex}, we can infer that $\Delta f$ is a function of $\alpha$ only, i.e., $\Delta f = \Delta f(\alpha)$. If there is a non-trivial solution (i.e., non-constant) for $\Delta f$ and $\Delta \psi$, it follows that the evolution equation must have the form of $\dot{\alpha}=\dot{\alpha}(\alpha)$, which is contradictory to Eq.~\eqref{eq: case1 evo alpha}. Hence $\Delta f$ and $\Delta \psi$ must both be constant. Since we fix the zeros of $f^{(i)}$ and $\psi^{(i)}$, $\Delta f$ and $\Delta \psi$ must be both zero, which implies that the free energy and dissipation potential densities are unique, and so is the dual dissipation potential.

\section{Further details on example 2} \label{appx: case 2}
\subsection{IB-VONNs with GMM} \label{Appx: case 2 IB_GMM} 

\begin{figure}[H]
  \centering
  \includegraphics[width=0.95\textwidth]{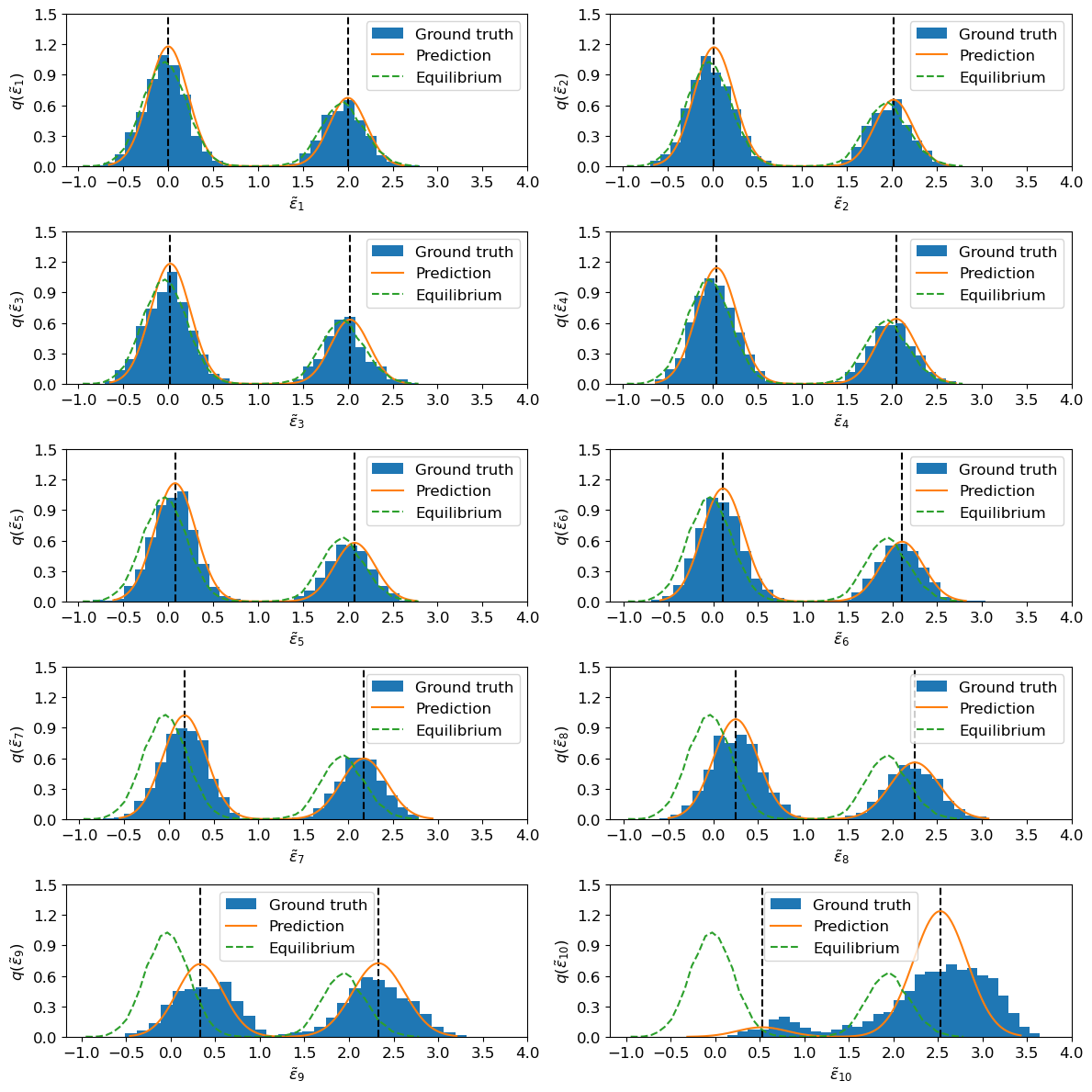}
  \caption{Microscopic marginal distributions $q(\tilde{\varepsilon}_i(t)|\boldsymbol{z}(t))$ computed by the reconstruction network with initial condition $L_{tot}=0.17$ (test set) at $t=1.0$. The histograms are computed from data, the solid lines represent the reconstructed probability distributions by IB-VONNs assuming a GMM for the decoder, and the dashed lines are the equilibrium marginal distributions. The vertical dashed lines show the modes of the GMM.}
  \label{fig: IB_GMM_pdf}
\end{figure}

\begin{figure}[H]
     \begin{subfigure}{0.44\textwidth}
         \centering
         \includegraphics[width=\textwidth]{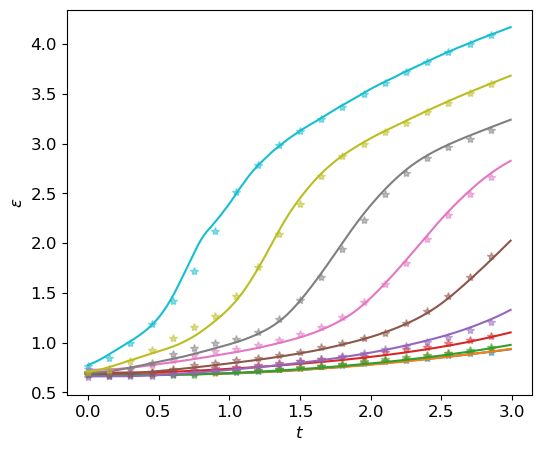}
         \caption{}
         \label{fig: IB_GMM_eps_Fex_input}
     \end{subfigure}
     \begin{subfigure}{0.54\textwidth}
         \centering
         \includegraphics[width=\textwidth]{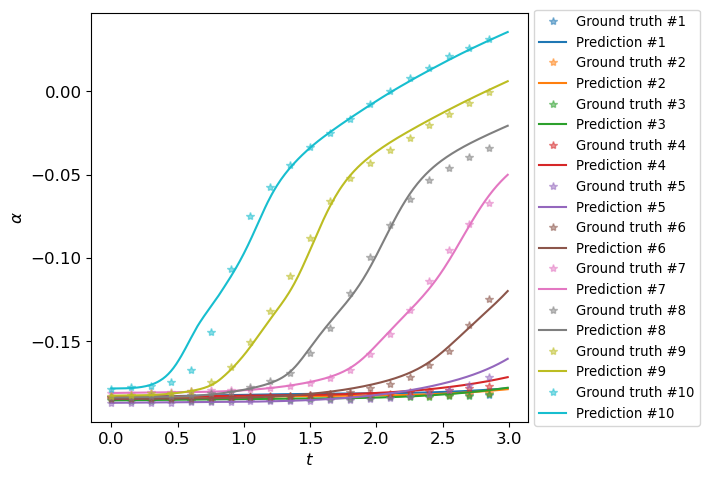}
         \caption{}
         \label{fig: IB_GMM_alpha_Fex_input}
     \end{subfigure}
\caption{Prediction of the evolution of (a) the mean strain and (b) the internal variable with the IB-VONNs framework with GMM on data with initial condition $L_{tot}=0.17$ (test set) with $F_{ex}$ as input. The stars represent the ground truth computed from Langevin simulations, and the solid lines capture the predictions by VONNs. The numbers in the legend are used to index the springs in the mass-spring chain of example 2.}
\label{fig: IB_GMM_eps_alpha_Fex_input}
\end{figure}

\begin{figure}[H]
     \centering
     \begin{subfigure}{0.40\textwidth}
         \centering
         \includegraphics[width=\textwidth]{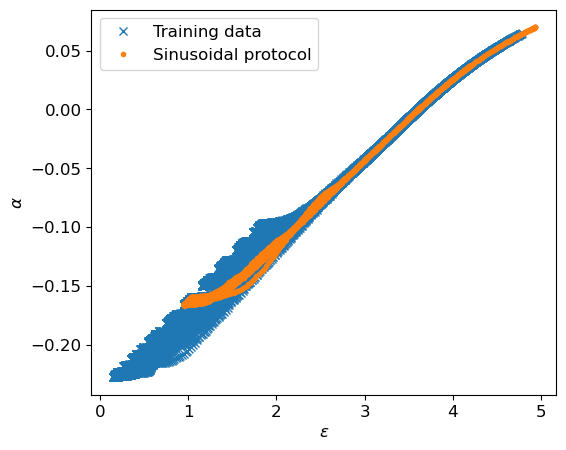}
         \caption{}
         \label{fig: IB_GMM_phase_space}
     \end{subfigure}
     \begin{subfigure}{0.54\textwidth}
         \centering
         \includegraphics[width=\textwidth]{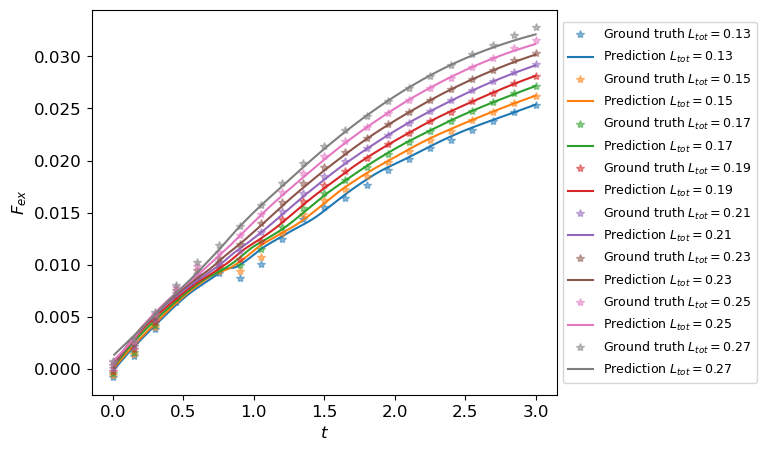}
         \caption{}
         \label{fig: IB_GMM_Fex}
     \end{subfigure}
     
     \centering
     \begin{subfigure}{0.43\textwidth}
         \centering
         \includegraphics[width=\textwidth]{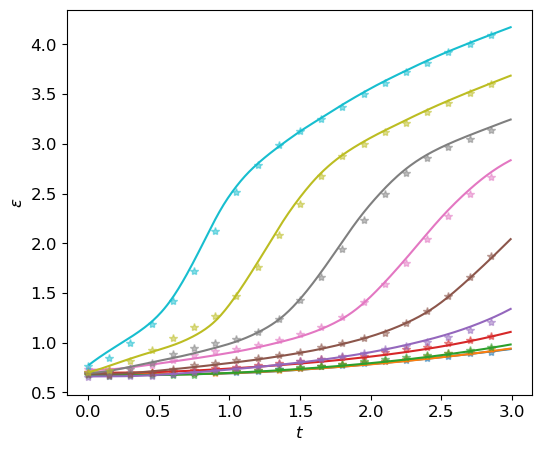}
         \caption{}
         \label{fig: IB_GMM_eps_v_input}
     \end{subfigure}
     \begin{subfigure}{0.54\textwidth}
         \centering
         \includegraphics[width=\textwidth]{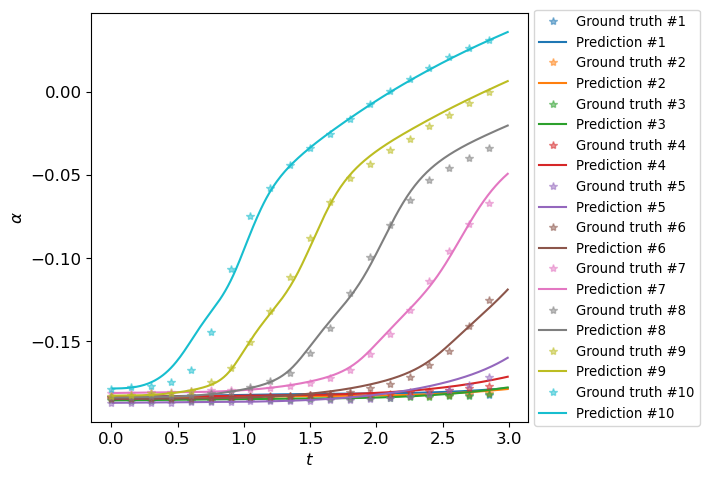}
         \caption{}
         \label{fig: IB_GMM_alpha_v_input}
     \end{subfigure}
\caption{Results from the IB-VONNs method with GMM. (a) The distribution of training data and the data generated by sinusoidal pulling protocol in the phase space of the state variables; (b) Predictions of the external force on the test set; Predictions of (c) the mean stain and (d) internal variable on data with initial condition $L_{tot}=0.17$ (test set). In (b) - (d), velocity of the last particle is fed to the model as input. The stars in (b) - (d) are ground truth computed from the Langevin simulations, and the solid lines are VONNs predictions. Numbers in the legends are the indices of the springs.}
\label{fig: IB_GMM_phase_Fex_eps_alpha_v_input}
\end{figure}

\begin{figure}[H]
     \centering
     \begin{subfigure}{0.43\textwidth}
         \centering
         \includegraphics[width=\textwidth]{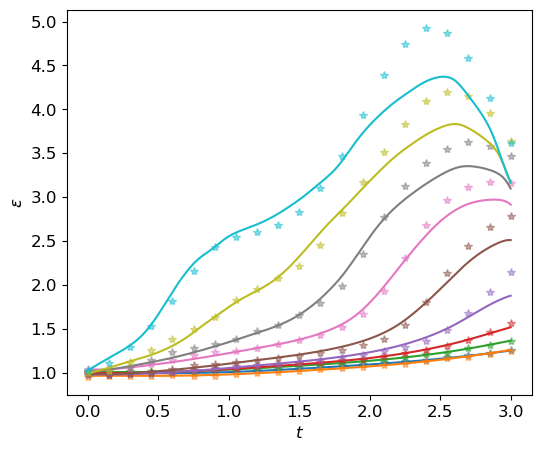}
         \caption{}
         \label{fig: IB_GMM_eps_v_input_sin}
     \end{subfigure}
     \begin{subfigure}{0.54\textwidth}
         \centering
         \includegraphics[width=\textwidth]{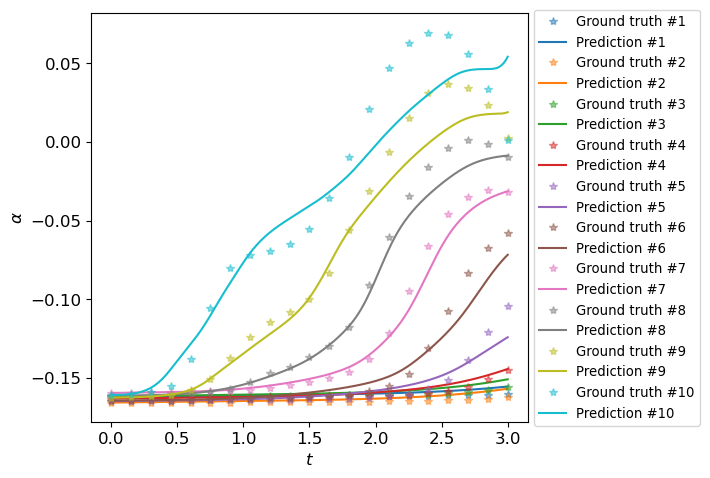}
         \caption{}
         \label{fig: IB_GMM_alpha_v_input_sin}
     \end{subfigure}
    \begin{subfigure}{0.54\textwidth}
         \centering
         \includegraphics[width=\textwidth]{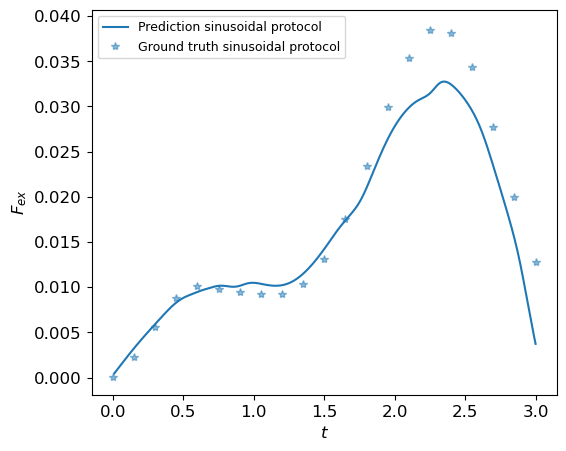}
         \caption{}
         \label{fig: IB_GMM_Fex_sin}
     \end{subfigure}
\caption{IB-VONNs with GMM tested on data with initial condition $L_{tot}=0.20$ and sinusoidal pulling protocol $\lambda(t)=vt(1 - e^t)(1+0.2\sin(\pi t))$. Predictions of (a) the mean strain, (b) the internal variable, and (c) external force with the velocity of the last particle as input. The stars are ground truth computed from Langevin simulations, and the solidlines are VONNs predictions. Numbers in the legends are the indices of the springs.}
\label{fig: IB_GMM_eps_alpha_v_input_sin}
\end{figure}

\subsection{IB-VONNs with CNFs} \label{Appx: case 2 IB_CNF}
\begin{figure}[H]
  \centering
  \includegraphics[width=0.95\textwidth]{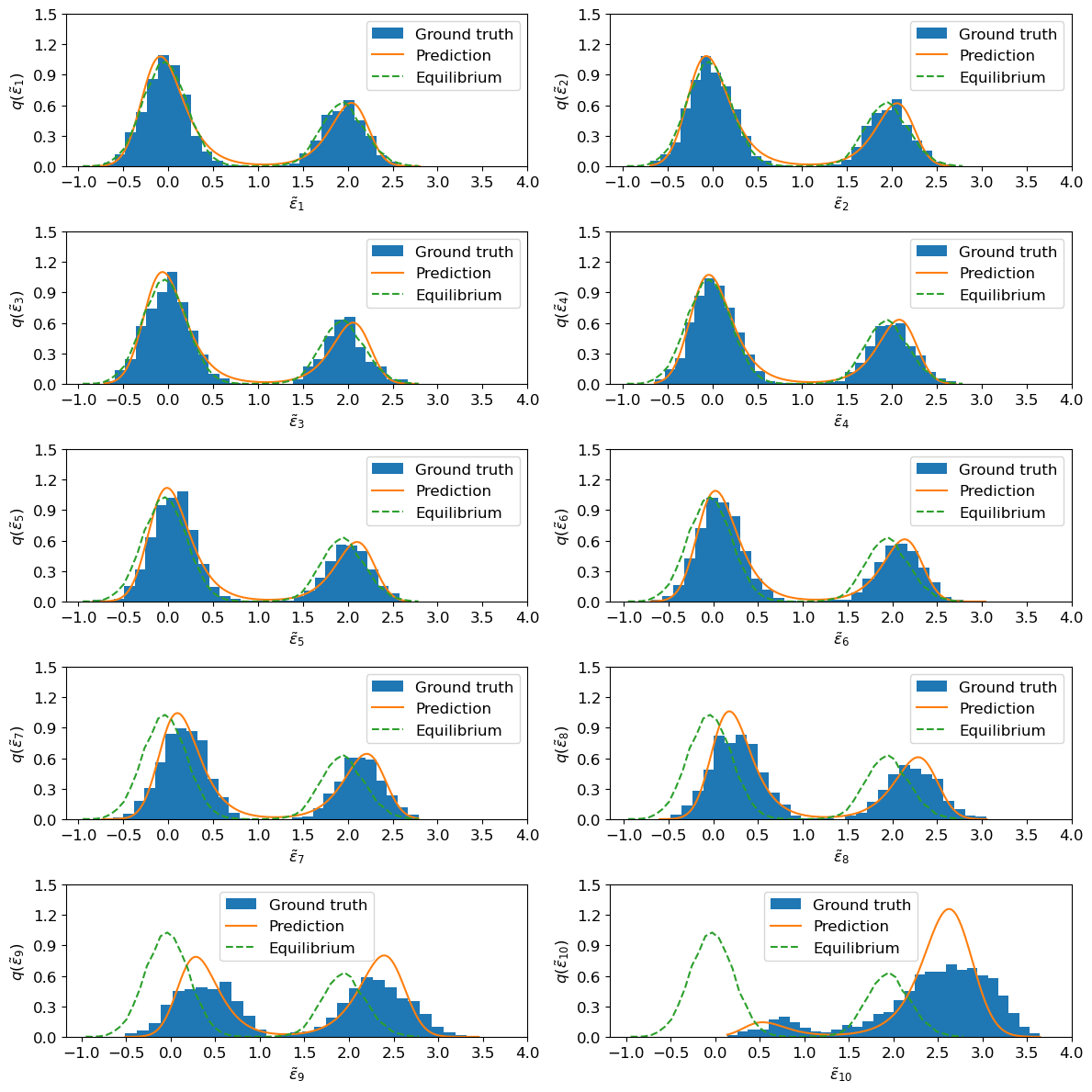}
  \caption{Microscopic distributions $q(\tilde{\varepsilon}_i(t)|\boldsymbol{z}(t))$ computed by IB-VONNs with CNFs with initial condition $L_{tot}=0.17$ (test set) at $t=1.0$. The histograms are computed from data, and the solid lines are the prediction of CNFs, and the dashed lines are the equilibrium marginal distributions.}
  \label{fig: IB_CNF_pdf}
\end{figure}

\begin{figure}[H]
     \begin{subfigure}{0.43\textwidth}
         \centering
         \includegraphics[width=\textwidth]{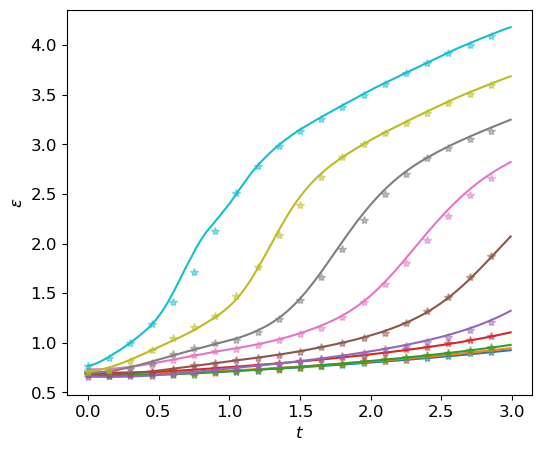}
         \caption{}
         \label{fig: IB_CNF_eps_Fex_input}
     \end{subfigure}
     \begin{subfigure}{0.54\textwidth}
         \centering
         \includegraphics[width=\textwidth]{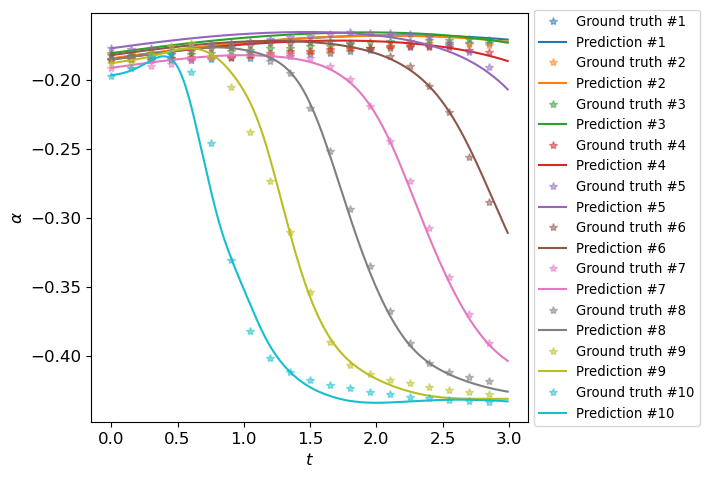}
         \caption{}
         \label{fig: IB_CNF_alpha_Fex_input}
     \end{subfigure}
\caption{Dynamics predictions of (a) the mean strain and (b) the internal variabl with the IB-VONNs framework with CNFs on data with initial condition $L_{tot}=0.17$ (test set) with $F_{ex}$ as input. The stars are ground truth computed from Langevin simulations, and the solid lines are VONNs predictions. Numbers in the legends are the indices of the springs.}
\label{fig: IB_CNF_eps_alpha_Fex_input}
\end{figure}

\printbibliography

\end{document}